\documentclass[12pt,a4paper]{iopart}

\usepackage[utf8]{inputenc}
\usepackage{graphicx}
\usepackage{color}
\usepackage{siunitx}
\usepackage{xspace}
\usepackage{overpic}
\usepackage{float}
\usepackage{amssymb}
\usepackage[numbers,sort&compress]{natbib}
\usepackage[all]{nowidow} 
\usepackage{soul}
\usepackage{hyperref}
\usepackage{amssymb}
\hypersetup{
	colorlinks=true,
	pdfstartview=FitV,
	linkcolor=blue,
	citecolor=blue,
	urlcolor=blue,
	pdfauthor={Hoppe Mathias, hoppe@chalmers.se},
	plainpages=false,
}

\newcommand{\bb}[1]{\textbf{#1}}
\newcommand{\dd}{\mathrm{d}}
\newcommand{\lambdac}{\lambda_{\mathrm{c}}}
\newcommand{\thetap}{\theta_{\mathrm{p}}}

\newcommand{\eqref}[1]{(\ref{#1})}

\definecolor{darkgreen}{RGB}{4,124,28}

\newcommand{\figlabel}[1]{\large{\textbf{\textsf{#1}}}}

\newcommand{\figlabelm}[1]{\textbf{\textsf{#1}}}
\newcommand{\figlabelsmall}[1]{\footnotesize{\textbf{\textsf{#1}}}}

\newcommand{\figlabelw}[1]{\large{\textbf{\textsf{\textcolor{white}{#1}}}}}

\begin{document}

    \title[Interpretation of runaway electron synchrotron and bremsstrahlung images]{Interpretation of runaway electron synchrotron and bremsstrahlung images}
    \author{
        M Hoppe$^{1}$,
        O Embr\'eus$^{1}$, 	
        C Paz-Soldan$^{2}$,
        RA Moyer$^{3}$,
    	T F\"ul\"op$^{1}$
    }
    
    \address{$^{1}$ Department of Physics, Chalmers University of Technology, G\"oteborg, Sweden}
    \address{$^{2}$ General Atomics, San Diego, California 92186, USA}
    \address{$^{3}$ University of California -- San Diego, La Jolla, California 92093, USA}
    \ead{hoppe@chalmers.se}
    
    \begin{abstract}
        The crescent spot shape observed in DIII-D runaway electron synchrotron radiation images is shown to result from the high degree of anisotropy in the emitted radiation, the finite spectral range of the camera and the distribution of runaways. The finite spectral camera range is found to be particularly important, as the radiation from the high-field side can be stronger by a factor $10^6$ than the radiation from the low-field side in DIII-D. By combining a kinetic model of the runaway dynamics with a synthetic synchrotron diagnostic we see that physical processes not described by the kinetic model (such as radial transport) are likely to be limiting the energy of the runaways. We show that a population of runaways with lower dominant energies and larger pitch-angles than those predicted by the kinetic model provide a better match to the synchrotron measurements. Using a new synthetic bremsstrahlung diagnostic we also simulate the view of the Gamma Ray Imager (GRI) diagnostic used at DIII-D to resolve the spatial distribution of runaway-generated bremsstrahlung.
    \end{abstract}
    
    \section{Introduction}\label{sec:intro}
     In the presence of a sufficiently strong electric field, thermal electrons can be accelerated to relativistic energies, and are commonly referred to as runaway electrons~\cite{Dreicer1959}. In tokamak disruptions, when the plasma current changes quickly and a strong electric field is induced, these electrons can carry a large fraction of the plasma current. If control of the plasma is then lost, the runaway electron beam will collide with the walls and can inflict severe damage. Therefore, the runaway electron phenomenon is regarded as among the greatest threats to future fusion reactors~\cite{Jayakumar1993,RosenbluthPutvinski1997,hollmann15iter,Hender2007,Boozer2017} and the performance of ITER relies on the successful mitigation of these relativistic particles.

Due to their rapid gyro-motion around magnetic field lines, runaway electrons will emit synchrotron radiation~\cite{Schwinger1949,Bekefi} almost entirely along their velocity vectors. The strong anisotropy of the radiation means that it contains information not only about the energy and radial distributions, but also about the pitch-angle distribution. Since it is emitted mainly at infra-red, and sometimes even visible wavelengths, synchrotron radiation is therefore an accessible diagnostic allowing indirect measurements of the runaway electron distribution function. The wealth of information contained in the synchrotron radiation also makes it an attractive candidate for benchmarking models of runaway dynamics.

Synchrotron radiation from runaway electrons was first studied on the TEXTOR tokamak~\cite{Finken1990} and has since been applied to many other tokamaks to study the runaway electron distribution function~\cite{Jaspers1995,Jaspers2001,Wongrach2014,Esposito2003,Esposito2017,Tinguely2015,Papp2016IAEA,Vlainic2015,Yu2013,Hollmann2013,PazSoldan2014,Zhou2013,Cheon2016,Tong2016}. Basic interpretation of the synchrotron radiation data obtained in experiments has been done ever since the first synchrotron radiation measurements, but in 1996 the first deeper analysis of the synchrotron radiation spot shape seen in camera images was carried out by Pankratov~\cite{Pankratov1996}, and later in 1999 the effects of the toroidal geometry on the synchrotron radiation spectrum were also considered~\cite{Pankratov1999}. The next major step in the modeling of synchrotron radiation came with \cite{Stahl2013} in 2013, where the validity of the asymptotic formulas given in \cite{Pankratov1999} in DIII-D and ITER were analyzed, and the importance of taking the full runaway electron distribution function into account was pointed out. In 2014, the theory derived in both \cite{Pankratov1996} and~\cite{Pankratov1999} was applied to the EAST tokamak in \cite{Zhou2014}, and analysis of the spot shape dependence on pitch angle and safety factor was done. Recently, more advanced synthetic synchrotron diagnostics have been developed that take both camera and magnetic field geometry into account. These are the \emph{Kinetic Orbit Runaway electron Code} (KORC)~\cite{Gomez2017KORC,Gomez2017}, which follows the runaway electron particle orbits and calculates the associated synchrotron radiation emission, and the \emph{Synchrotron-detecting Orbit Following Toolkit} (SOFT)~\cite{Hoppe2017} which utilizes the guiding-center approximation to rapidly calculate the synchrotron radiation. In \cite{HoppeMSc,Hoppe2017}, SOFT was applied to a specific Alcator C-Mod scenario in order to discern the effect of, among others, the runaway electron energy and radial profile, and it was found that both are crucial for interpreting the synchrotron radiation spot.

Runaway electron bremsstrahlung is very similar to synchrotron radiation in that it too is directed almost entirely along the electron's velocity vector. Thus, much of the theory derived in \cite{Hoppe2017} applies also to the study of bremsstrahlung and a synthetic diagnostic for runaway electron bremsstrahlung could hence be implemented similarly to synchrotron radiation in SOFT. In this paper, we have extended SOFT with bremsstrahlung capabilities and will use it to simulate the \emph{Gamma Ray Imager} (GRI) diagnostic~\cite{Pace2016,Cooper2016}, situated at DIII-D. The GRI consists of a lead pinhole camera and an array of gamma-ray detectors and therefore provides both spatial and spectral resolution of runaway electron bremsstrahlung emission. The large amount of information provided by the GRI gives a unique view into the evolution of runaway electrons during experimental scenarios, and may be able to provide sufficient data for the first robust calculations of the runaway electron distribution function from runaway electron radiation measurements.

In this paper we investigate the leading causes behind the particular shape of a DIII-D synchrotron radiation image, and compare the distribution function predicted by a kinetic model for runaways to experimental synchrotron and bremsstrahlung measurements for the same experimental discharge. With a qualitative model and simulations we identify the sources of various features in the image and explain the characteristic crescent spot shape of the DIII-D synchrotron image. By solving the Fokker-Planck equation for runaways numerically we predict a distribution function for the runaways in the investigated DIII-D scenario in order to assess the validity of the model. We also present the first SOFT simulations of the GRI using which we discuss similarities and differences between observed runaway synchrotron and bremsstrahlung radiation.

In Section~\ref{sec:model}, a qualitative model is presented and we identify the most important quantities affecting a synchrotron radiation image to be the so called surface-of-visibility, the finite spectral range of the camera used and the distribution function. In Section~\ref{sec:experiment}, kinetic simulations of the runaway electron distribution function in combination with SOFT simulations are conducted for DIII-D discharge 165826. The disagreement of these simulations with experiment is related to possibly missing physics in the kinetic model used. The paper concludes with a discussion in Section~\ref{sec:conclusions} about what can be said about runaway electron radiation investigations in low-density scenarios similar to DIII-D discharge 165826.

    \section{Qualitative model of synchrotron radiation}\label{sec:model}
	Synchrotron radiation is emitted by highly relativistic charged particles in magnetic fields due to their cyclotron motion~\cite{Schwinger1949,Bekefi}, and characteristic for this type of emission is the strong forward beaming of radiation along the particle's velocity vector. Because of this, electrons will only be visible to the observer in the regions of the tokamak where the electrons move directly towards the detector, and synchrotron radiation images therefore typically show synchrotron radiation ``spots'' appearing on only one side of the tokamak. The angular spread of the emission is $\sim\gamma^{-1}$ (considering the average over all emitted wavelengths), where $\gamma$ is the relativistic gamma factor, which for high energy runaway electrons means that the radiation can be approximated as emitted almost exactly on a straight line along the electron's velocity vector. The model in which this approximation is made, henceforth referred to as the ``cone model'' for reasons  soon to be explained, was implemented in the SOFT code~\cite{Hoppe2017} and compared to a model taking the full angular spread of the radiation into account. The comparison showed that these two models were in good agreement with each other. The simulations conducted in this paper will all use the cone model due to its superior computational performance.

The name of the cone model stems from the fact that the guiding-center of an electron can be seen as emitting a cone of radiation with opening angle $\thetap = \arccos(v_\parallel / v)$ around its velocity vector. It is assumed that all radiation is emitted uniformly across this cone, and in the cone model used in this paper it is further assumed that the synchrotron power received in a wavelength interval $[\lambda, \lambda+\dd\lambda]$ is given by~\cite{Bekefi}
\begin{equation}
    \frac{\dd P}{\dd\lambda} = \frac{1}{\sqrt{3}}\frac{ce^2}{\epsilon_0\lambda^3\gamma^2} \int_{\lambdac/\lambda}^\infty K_{5/3}(l)\,\dd l.\label{eq:specdist}
\end{equation}
Here, $c$ is the speed of light, $e$ is the elementary charge, $\epsilon_0$ is the permittivity of free space, $\gamma$ is the Lorentz factor for the electron, $\lambdac = 4\pi m_ec\gamma_\parallel / 3\gamma^2 e B$, $m_e$ is the electron mass, $\gamma_\parallel = (1-v_{\parallel})^{-1/2}$, $v_\parallel$ is the electron's speed along the magnetic field, $B$ is the magnetic field strength and $K_{5/3}$ is a modified Bessel function of the second kind.

When analyzing synchrotron radiation from runaway electrons, it turns out that the appearance of synchrotron radiation spots as seen by a camera is determined mainly by three different effects:
\begin{itemize}
    \item The so called \textbf{surface-of-visibility} (SOV), to be explained in Section~\ref{sec:sov}, which is a geometric feature and results from the anisotropy of the synchrotron radiation emission.
    \item The \textbf{amount of radiation emitted} and, more importantly, detected. The location of the wavelength interval that a synchrotron camera operates in can significantly alter the appearance of a synchrotron spot depending on how close it is to the peak wavelength of the radiation received by the camera.
    \item The momentum-space \textbf{distribution function} of runaway electrons, the shape of which determines which particles will dominate the emission.
\end{itemize}
In what follows we will describe the importance of each of these effects and show how a synchrotron spot can be built up successively by them.

\subsection{Surface-of-visibility}\label{sec:sov}
\begin{figure}
    \centering
    \begin{overpic}[width=0.24\textwidth]{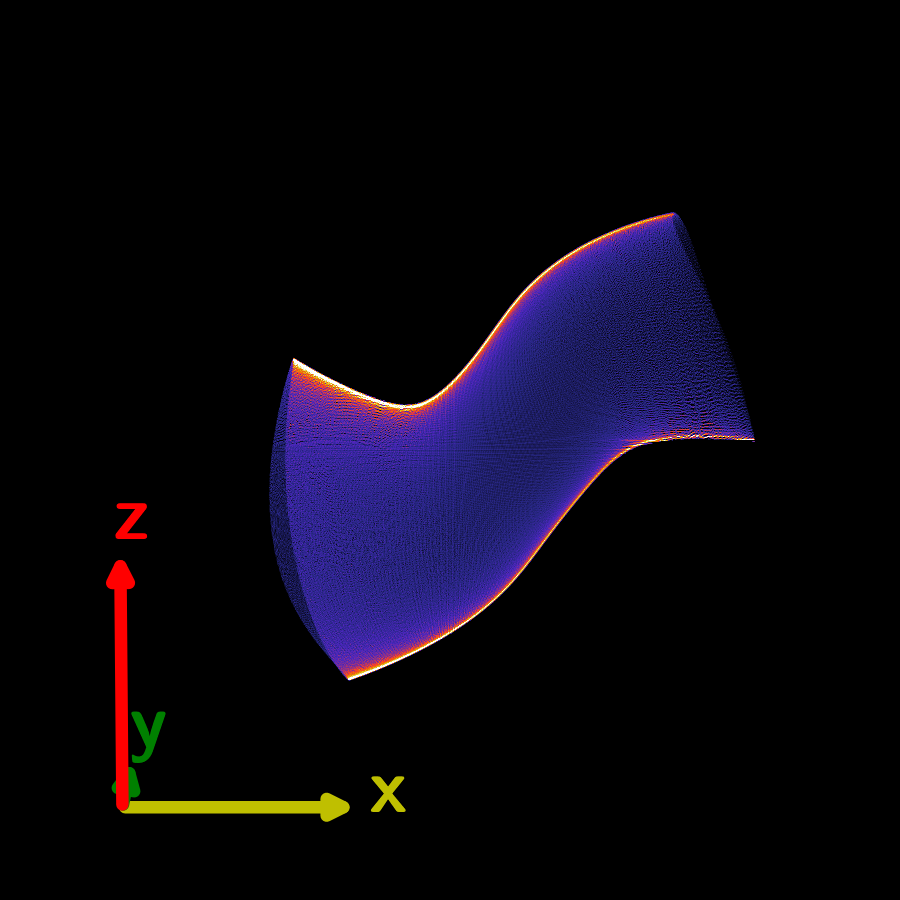}
        \put(5,85){\figlabelw{(a)}}
    \end{overpic}
    \begin{overpic}[width=0.24\textwidth]{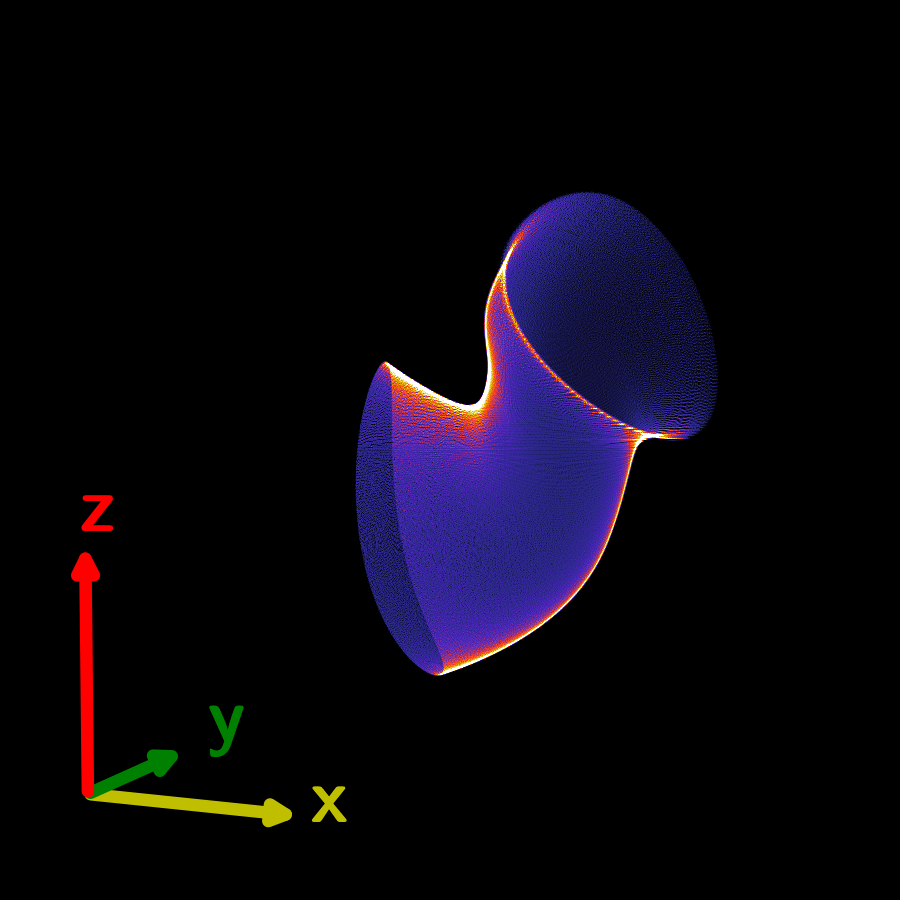}
        \put(5,85){\figlabelw{(b)}}
    \end{overpic}
    \begin{overpic}[width=0.24\textwidth]{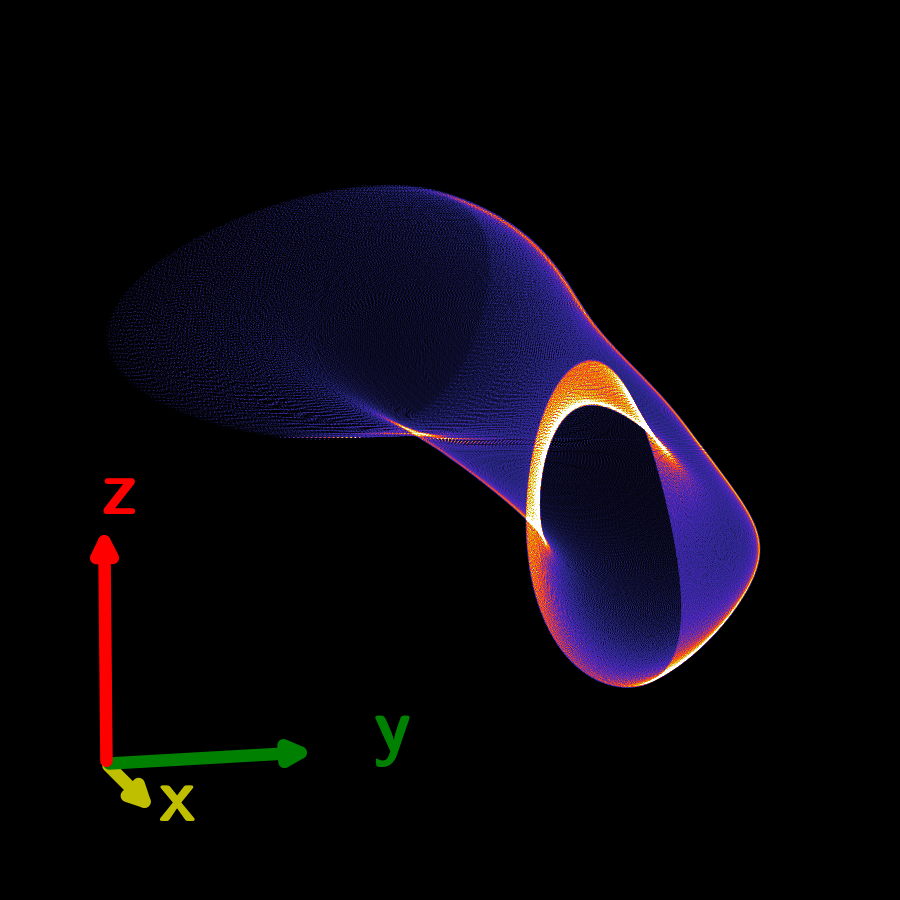}
        \put(5,85){\figlabelw{(c)}}
    \end{overpic}
    \begin{overpic}[width=0.24\textwidth]{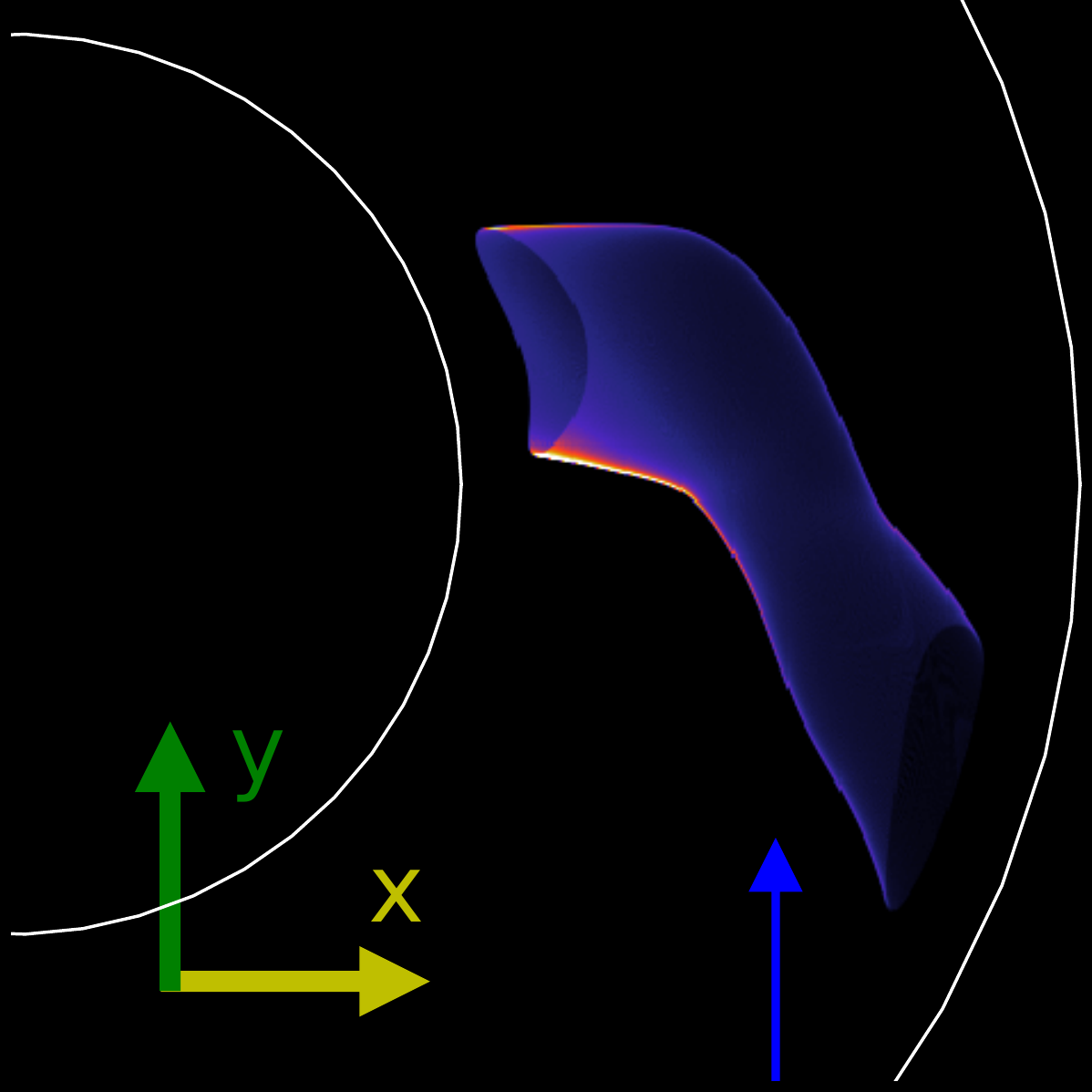}
        \put(5,85){\figlabelw{(d)}}
    \end{overpic}
    \caption{Different views of the surface-of-visibility corresponding to runaways with $\thetap = \SI{0.16}{rad}$ in DIII-D. Between each of the figures (a)-(c) the SOV is rotated clockwise, and in (d) a top-down view of the tokamak and SOV is shown. The blue arrow in (d) indicates the location and direction of the camera.}
    \label{fig:sov}
\end{figure}

Aside from bringing the computational advantages of not having to resolve the full gyro-orbit, the cone model also provides a simple framework for qualitative reasoning about the region from which synchrotron radiation can be detected. Since synchrotron radiation is only detected when the emitting particle is moving directly towards the detector, the overall shape of the observed synchrotron spot is mainly determined by the magnetic field geometry~\cite{Pankratov1996,Zhou2014}. Neglecting drift velocities, the cone model gives a very simple condition for a particle to be seen by the detector:
\begin{equation}\label{eq:sov}
    \left|\hat{\bb{b}}(\bb{x})\cdot\hat{\bb{r}}(\bb{x})\right| = |\cos\thetap|.
\end{equation}
Here $\hat{\bb{b}}(\bb{x})$ is the magnetic field unit vector in the point $\bb{x}$ in space, $\hat{\bb{r}} = (\bb{x}-\bb{x}_0) / |\bb{x}-\bb{x}_0|$ is the unit vector pointing from the detector, located at $\bb{x}_0$, to $\bb{x}$ and $\thetap$ is the pitch angle at $\bb{x}$ under consideration. For a fixed value of $\thetap$, the solution to this equation is a surface in real space, which we call the \emph{surface-of-visibility} (SOV). It is clear from Eq.\ \eqref{eq:sov} that aside from a strong dependence on the magnetic field and the particle's pitch angle, the shape of the SOV is also strongly dependent of the placement of the detector. In Figure~\ref{fig:sov}, different projections of the surface-of-visibility for a given $\bb{x}_0$ lying in the midplane and runaway electron population with $\thetap = \SI{0.16}{rad}$ in DIII-D is shown, revealing its cylindrical structure. Note that the surface-of-visibility in Fig.~\ref{fig:sov} is the surface corresponding to particles given an initial pitch angle $\thetap = \SI{0.16}{rad}$ in the outer midplane. The pitch angle then varies together with the magnetic along the particle orbit.

The importance of the magnetic field geometry for determining the synchrotron spot shape was emphasized already by Pankratov in 1996~\cite{Pankratov1996}, but the realization that what is perceived as a ``spot'' in synchrotron images is in fact the projection of a \emph{surface} of finite toroidal extent could be even more important when considering momentum-space distributed populations of runaway electrons. The reason for this is that SOVs for particles with small pitch angles (typically $\thetap\lesssim \SI{0.20}{rad}$) close on themselves, and the line-integrated contribution from a line-of-sight passing through the edge of such a surface will be greater due to that the line-of-sight tangents the surface. The edges of the projected SOV therefore tend to be significantly brighter than other parts of the SOV, as is exemplified in Fig.~\ref{fig:sov}. Because of this the edges of single-particle synchrotron spots (i.\,e.\ spots corresponding to a specific set of pitch angle $\thetap$ and momentum $p$) tend to dominate synchrotron spots from momentum-space distributed runaway electron populations and ``fill in'' different parts of the overall spot.

\subsection{Synchrotron emission and camera spectral range}\label{sec:spectrumeffect}
The amount of radiation emitted by a particle is another important quantity affecting the appearance of a synchrotron spot. From the $P\propto p_\perp^2 B^2\propto p^2B^3$ scaling of synchrotron emission, where $p$ is momentum and $p_\perp = p\sin\thetap\propto p\sqrt{B}$ is the component of $p$ perpendicular to the magnetic field $\bb{B}$, we expect a particle to emit more radiation when it passes through the high-field side of a tokamak, both because the magnetic field is stronger there, and because of the larger pitch angle the particle will have due to the stronger magnetic field. It turns out however, that while the synchrotron emission from the high-field side is typically stronger than that from the low-field side, the amount of synchrotron radiation received by a camera from the high-field side can scale much more strongly with magnetic field than $B^3$. In fact, as we will now show, the ratio between emission from the HFS to the emission from the LFS can be on the order of $10^6$ in DIII-D.

The reason for the strong scaling of the detected radiation stems from the finite spectral range of the camera. In most present-day tokamaks, the runaway electrons emit most of their radiation at wavelengths around a few micrometers, while visible light cameras seeing wavelengths up to around $\SI{900}{nm}$ are used. In this case it can be shown that the short-wavelength ($\lambda\ll\lambdac$) asymptotic expansion of Eq.~\eqref{eq:specdist} is
\begin{equation}
    \frac{\dd P}{\dd\lambda}\sim \exp\left( -\frac{\lambdac}{\lambda} \right).
\end{equation}
By introducing the critical radius
\begin{equation}
    R_\mathrm{c} = \frac{B_0R_0}{2m_e}\left( \frac{3\gamma e\lambda\sqrt{\mu}}{\pi c^2} \right)^{2/3},
\end{equation}
with $\mu = p_\perp^2 / 2m_eB$ being the magnetic moment, and assuming the magnetic field strength as a function of major radius to be $B(R) = B_0R_0/R$, where $B_0$ and $R_0$ are the magnetic field strength and major radial location respectively of the magnetic axis, we can write the detected synchrotron power as
\begin{equation}\label{eq:expscale}
    P\sim \exp\left[ -\left(\frac{R}{R_{\mathrm{c}}}\right)^{3/2} \right].
\end{equation}
We can also express the ratio between maximum emission of a particle (at the innermost point of its orbit) to its minimum emission (at the outermost point of its orbit) as
\begin{eqnarray}
    \frac{P_{\mathrm{max}}}{P_{\mathrm{min}}} &= \exp\left[ \frac{(R_0+\Delta R/2)^{3/2}-(R_0-\Delta R/2)^{3/2}}{R_{\mathrm{c}}^{3/2}} \right] \approx\\
    &\approx\exp\left[\frac{3}{2}\frac{\Delta R}{R_0}\left(\frac{R_0}{R_{\mathrm{c}}}\right)^{3/2} \right]
\end{eqnarray}
where $\Delta R$ is the distance between the inner- and outermost points of the particle orbit.
The critical radius, $R_\mathrm{c} = R_\mathrm{c}(\lambda)$, can be interpreted as the major radius at which $\lambdac = \lambda$, i.e. where the particle emits most of its radiation near the wavelength $\lambda$.

The exponential scaling with particle position Eq.~\eqref{eq:expscale} is the result of observing radiation in only a narrow range of wavelengths far from the wavelengths at which synchrotron emission peaks. In a typical DIII-D scenario, where $\lambda\approx\SI{900}{nm}$ and $\gamma\approx 30$, we find that $R_\mathrm{c}\approx 0.1 R_0$. A DIII-D runaway electron which moves a distance $\Delta R = \SI{1}{m}$ in major radius during its orbit therefore emits more on the high-field side compared to the low-field side by a factor $P_{\mathrm{max}}/P_{\mathrm{min}} \sim 10^6$, i.e.\ a six orders of magnitude difference. Had we instead been able to observe at a wavelength $\lambda$ closer to $\lambdac$, or even all radiation, the difference would merely have been about a factor of six. This shows that synchrotron radiation can appear significantly brighter in a region of space, without the number of runaways necessarily being higher in that region.

\subsection{Runaway electron distribution function}\label{sec:distfunc}
In many previous studies, synchrotron spectra and spot shapes from single particles have been used to model experimentally observed spectra~\cite{Jaspers2001,Yu2013} and spot shapes~\cite{Zhou2014}. As was shown in~\cite{Stahl2013} however, aside from the difficulties of interpreting the results (runaway electrons are rarely homogeneous enough for a single particle to satisfactorily approximate the population), taking the distribution function of runaway electrons into account can significantly alter the simulated spectra. For the synchrotron spot, the difference can be even more dramatic, as the brightest features of a number of single particle-spot shapes will come together and create an overall pattern which does not necessarily resemble any of the single particle spot shapes.

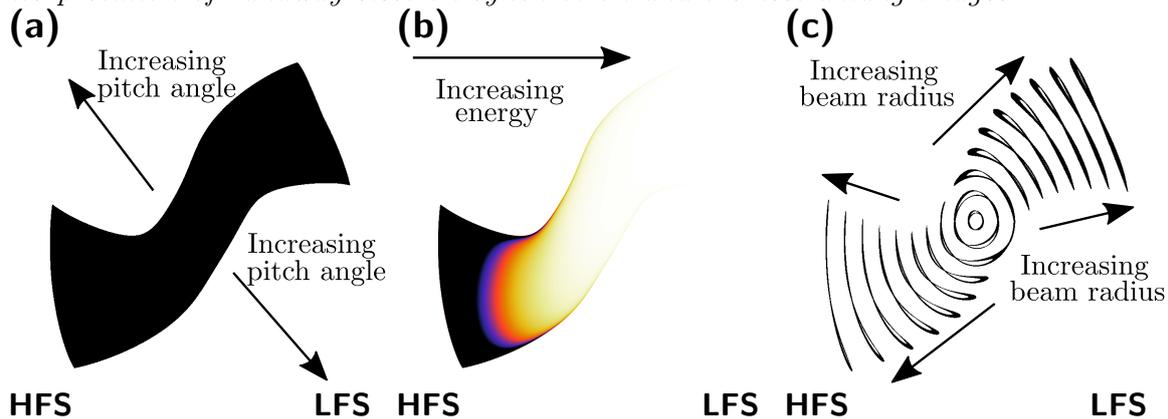
\begin{figure}
    \centering
    \begin{overpic}[width=0.32\textwidth]{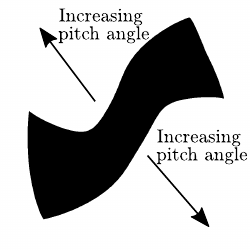}
        \put(0,100){\figlabel{(a)}}
        \put(0,0){\figlabelm{HFS}}
        \put(80,0){\figlabelm{LFS}}
    \end{overpic}
    \begin{overpic}[width=0.32\textwidth]{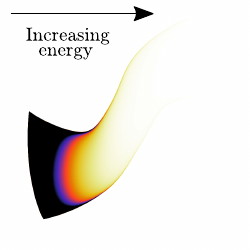}
        \put(0,100){\figlabel{(b)}}
        \put(0,0){\figlabelm{HFS}}
        \put(80,0){\figlabelm{LFS}}
    \end{overpic}
    \begin{overpic}[width=0.32\textwidth]{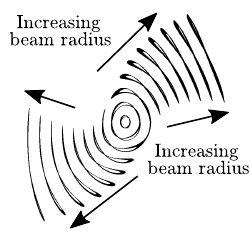}
        \put(0,100){\figlabel{(c)}}
        \put(0,0){\figlabelm{HFS}}
        \put(80,0){\figlabelm{LFS}}
    \end{overpic}
    \caption{An illustration of how the three parameters of the runaway electron distribution function affect the shape of the synchrotron spot. The synchrotron radiation spot appears on the right side of the tokamak, i.e. to the right of the axis of symmetry. \textbf{(a)} A larger pitch angle implies greater vertical extent of the synchrotron spot and mainly affects the SOV. \textbf{(b)} Due to the finite spectral range effect of Section~\ref{sec:spectrumeffect}, increasing the runaway electron energy causes radiation intensity to increase on the LFS relative to the HFS. \textbf{(c)} A larger runaway electron beam causes the synchrotron spot to be larger in the horizontal direction. Each of the slices of the spot in (c) comes from particles initiated at one radius, and they converge towards the magnetic axis in the center of the spot.}
    \label{fig:spoteffects}
\end{figure}

The pitch angle and energy of a particle, together with the flux surface on which the particle moves (radial position), together determine the shape of the SOV and at which wavelength most of the synchrotron radiation is emitted. In Fig.~\ref{fig:spoteffects} the effect of each of these parameters on the synchrotron spot is sketched and the arrows indicate how the spot shape changes with increasing pitch angle, energy and radial location. Fig.~\ref{fig:spoteffects}a indicates that the pitch angle mainly determines the vertical extent of the SOV, while the radial location of the particle mainly affects the horizontal extent of the SOV, as shown in Fig.~\ref{fig:spoteffects}c. The particle energy is mainly tied to the finite spectral range effect described in Section~\ref{sec:spectrumeffect}, as is illustrated in Fig.~\ref{fig:spoteffects}b. As the particle energy is increased, the peak of the emitted spectrum increases as well, meaning that the critical radius $R_{\mathrm{c}}$ also increases. For the synchrotron radiation spot this means that the low-field side part of the spot (right side of image) gains intensity relative to the high-field side part (left side of the image) giving a more even distribution of radiation horizontally across the SOV. A more in-depth description of how the spot shape varies with different parameters can be found in~\cite{Hoppe2017,HoppeMSc}.

The synchrotron spot of a certain population of runaway electrons will be the weighted average of several synchrotron spots, each corresponding to a unique set of runaway electron energy, pitch angle and radial location, i.e.\ individual particles. The weight is the distribution function which determines the relative importance of different particles in accordance with how likely they are to be found in the population and determines the overall spot shape.

A useful quantity that provides much information about the radiation from a distribution of runaway electrons is the density of radiation in momentum space,
\begin{equation}\label{eq:radiationdensity}
    F(p_\parallel, p_\perp) = \hat{I}(p_\parallel, p_\perp)f(p_\parallel, p_\perp),
\end{equation}
where $f(p_\parallel, p_\perp)$ is the runaway electron distribution function and $\hat{I}(p_\parallel, p_\perp)$ is the amount of radiation emitted by a particle with the given momentum, henceforth referred to as \emph{weight function} (different for synchrotron or bremsstrahlung). Due to the $p_\perp^2$ scaling (or even stronger, as explained above) of the synchrotron emission, i.e.\ $\hat{I}_s$, the most numerous particle type is not necessarily the one that emits the most synchrotron radiation. While $\hat{I}$ tends to grow monotonically with momentum $p$, and $f$ tends to decrease with $p_\parallel$ and $p_\perp$, the radiation density $F$ generally has a global maximum in momentum space different from $p_\parallel = p_\perp = 0$. The maximum of $F$ can be considered a ``super''-particle which will dominate emission of a particular radiation type from a given distribution of runaway electrons. Often, the single particle spectrum that best matches the distribution averaged spectrum is that of the super particle, and the observed spot shape from the distribution has often many similarities to the spot shape of the super particle. It should be noted though that the only significance of the super particle is that it is the maximum of $F$. It is not (necessarily) the particle with the highest energy or pitch angle, and no general statement about the most common electron momentum can be made.

\subsection{Similarities to runaway bremsstrahlung emission}
Just like synchrotron radiation, bremsstrahlung is highly anisotropic and directed mainly in the particle's direction of motion for highly relativistic particles. The average angular spread of bremsstrahlung is the same $\gamma^{-1}$ as for synchrotron radiation, which means that bremsstrahlung also gives rise to a surface-of-visibility as discussed above, that is the same as that of synchrotron radiation. The cone model used in SOFT for synchrotron radiation can thus also be used for bremsstrahlung, but with the formula for received synchrotron radiation power replaced with the formula for the number of bremsstrahlung photons $\dd N_\gamma$ emitted per unit photon energy $\dd k$ (using the Born-approximation cross-section for a fully ionized plasma~\cite{Koch1959} summed over all ion species)
\begin{eqnarray}\label{eq:bremsspec}
     \frac{\dd N_\gamma}{\dd k} &= \frac{\alpha n_e Z_\mathrm{eff} e^4 v p'}{k p}\Biggr\{
    \frac{4}{3} - 2\gamma'\gamma\frac{p^2+p'^2}{p'^2p^2} + \epsilon\frac{\gamma'}{p^3}+\epsilon'\frac{\gamma}{p'^3} - \frac{\epsilon\epsilon'}{p'p}+L\Biggr[ \frac{8}{3}\frac{\gamma'\gamma}{p'p}  \nonumber \\
    &+k^2\frac{\gamma'^2\gamma^2+p'^2p^2}{p'^3p^3} + \frac{k}{2p'p}\left( \epsilon\frac{\gamma'\gamma+p^2}{p^3} - \epsilon'\frac{\gamma'\gamma+p'^2}{p'^3} + 2k\frac{\gamma'\gamma}{p'^2p^2}  \right) \Biggr] 
    \Biggr\}, \nonumber\\\
     \epsilon &= \ln\frac{\gamma+p}{\gamma-p} \\
     \epsilon' &= \ln\frac{\gamma'+p'}{\gamma'-p'} \nonumber \\
     L &= 2\ln\frac{\gamma'\gamma+p'p-1}{k}, \nonumber
\end{eqnarray}
where $\alpha=e^2/4\pi\varepsilon_0 \hbar c \approx 1/137$ is the fine-structure constant, the photon energy $k$ is defined in units of $m_e c^2$, and the normalized ingoing and outgoing electron momenta $p=\gamma v/c$ and $p'$ are related through $p'=\sqrt{\gamma'^2-1}$, and $\gamma' = \gamma-k$.

Measurements of bremsstrahlung from runaway electrons is today a standard diagnostic at most larger tokamak experiments, and the data acquired is often used to study the temporal evolution of runaway electrons and sometimes even to measure the runaway electron energy distribution function~\cite{PazSoldan2017,Shevelev2017}. Bremsstrahlung from runaways has also previously been modeled, for example in the Tore Supra tokamak~\cite{Peysson2008,Nilsson2012}. At DIII-D, a novel technique for measuring not only the bremsstrahlung spectrum, but also its spatial distribution, has been developed and is called the \emph{Gamma-Ray Imager} (GRI)~\cite{Pace2016,Cooper2016}. The GRI combines a lead pinhole camera with gamma-ray detectors, thus functioning as a camera for gamma rays with energies in the range 1-$\SI{60}{MeV}$. From a theoretical point-of-view, the GRI is an ideal tool for studying runaways since the spectrum is also measured by each gamma-ray detector, thus providing a set of images at different photon-energies rather than just one single image. The weak pitch angle dependence in the bremsstrahlung emission is also advantageous, as it avoids the finite spectral range effect experienced with synchrotron radiation, which tends to obscure large parts of the spatial information.

    \section{Runaway electron radiation images in DIII-D}\label{sec:experiment}
    \begin{figure}
    \centering
    \begin{minipage}{0.49\textwidth}
        \centering
        \begin{overpic}[width=2.5in]{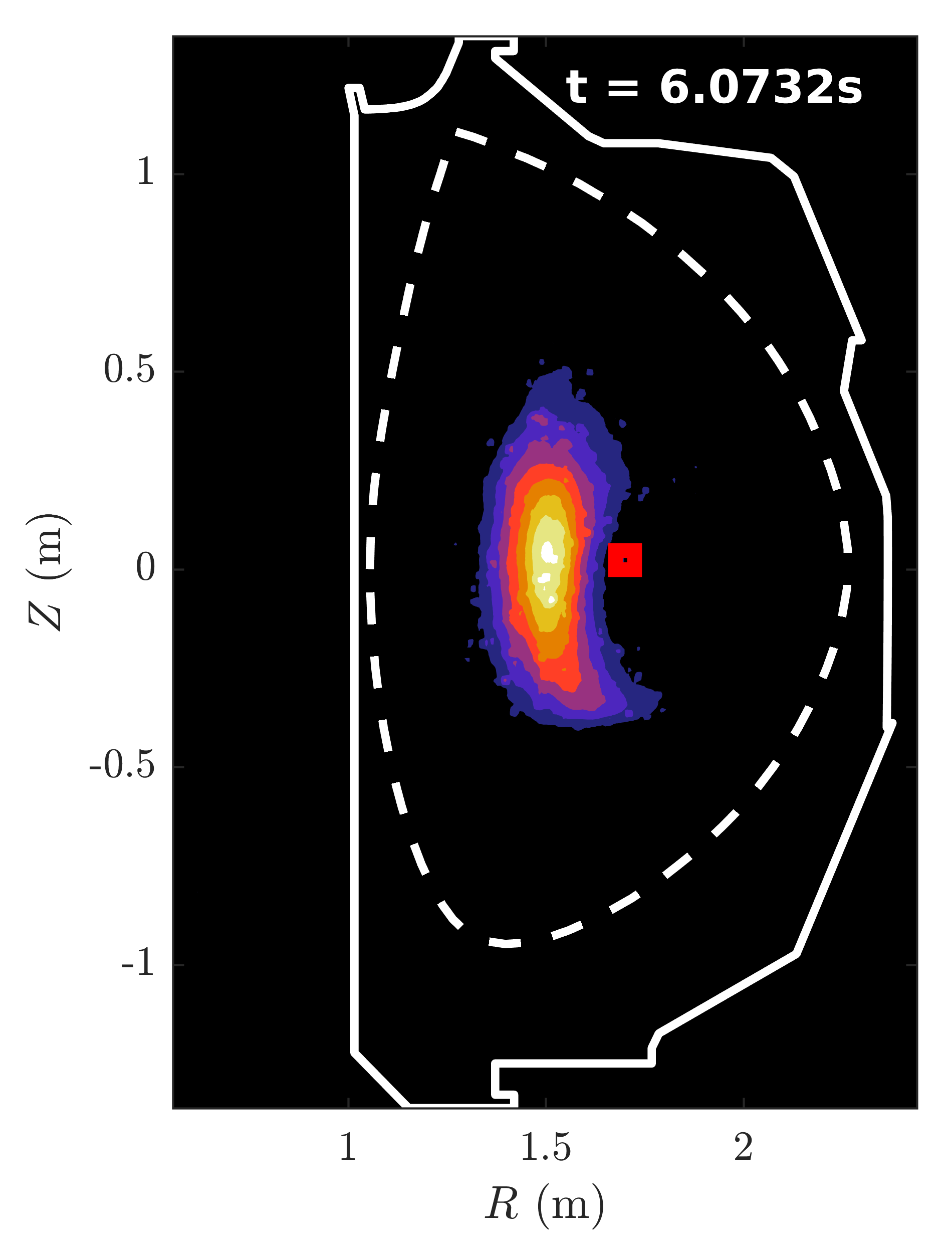}
            \put(16,90){\figlabelw{(a)}}
        \end{overpic}
    \end{minipage}
    \begin{minipage}{0.49\textwidth}
        \centering
        \begin{overpic}[width=2.5in]{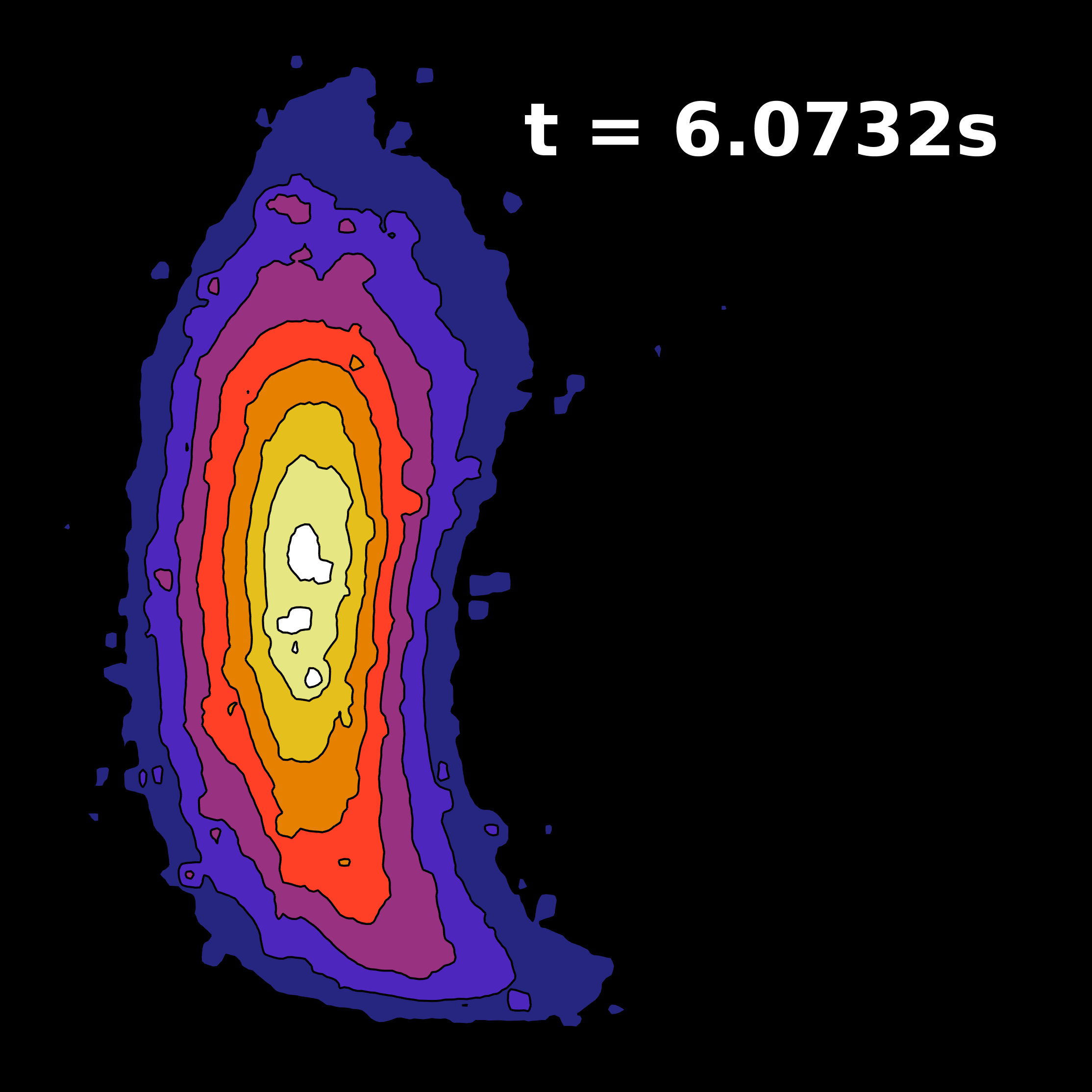}
            \put(4,90){\figlabelw{(b)}}
        \end{overpic}
    \end{minipage}
    \caption{Synchrotron radiation from runaway electrons observed at $t = \SI{6.0732}{s}$ during DIII-D discharge 165826. \textbf{(a)} Synchrotron radiation mapped to the tangency plane with a wall cross-section and separatrix curve overlaid. The red marker indicates the location of the magnetic axis. \textbf{(b)} Synchrotron radiation in the pixel plane.}
    \label{fig:synchcamera}
\end{figure}

Comparing and validating models of runaway electron dynamics against experiments is of crucial importance in order for any confidence to put in the models. The strong dependence on the distribution function seen in both synchrotron and bremsstrahlung emission, as discussed in the previous section, makes both types of radiation attractive diagnostics for this purpose. In this section we start by solving the spatially homogeneous kinetic equation numerically, taking plasma parameters from a DIII-D discharge as input, and use SOFT to compute the corresponding synchrotron and bremsstrahlung images. As the synthetic synchrotron images are found to disagree with the experimental images, we assess the properties required by the distribution function for agreement. We conclude the section with an analysis of bremsstrahlung images, which we compare to experimental data and discuss similarities and differences to synchrotron images.

We will analyze DIII-D discharge 165826~\cite{PazSoldan2017}, a quiescent flattop runaway discharge~\cite{PazSoldan2014} which is carried out in two phases. In the first, low-density phase the runaway electron population is steadily built up through mainly primary (Dreicer) generation. When the runaway electrons have reached a critical density, nitrogen and deuterium is injected to initiate the dissipation phase, during which primary runaway electron generation ceases and effects such as avalanche generation and synchrotron/bremsstrahlung damping play a key role in the evolution of the runaway electron distribution function.

The fast synchrotron camera diagnostic used during the discharge was directed tangentially towards the plasma and detected all radiation emitted in a narrow band near wavelength $\SI{890}{nm}$. It shows a characteristic crescent synchrotron radiation spot shape, mainly originating from the HFS, with a maximum that is approximately vertically aligned with the magnetic axis, similar to what has been observed in other DIII-D low-density discharges~\cite{PazSoldan2014}.

For the following discussion we pick the synchrotron image corresponding to $t = \SI{6.0732}{s}$, which is shown in Fig.~\ref{fig:synchcamera} and is representative for the discharge. The image reveals that most of the synchrotron radiation is seen on the HFS, which based on our discussion in Section~\ref{sec:spectrumeffect} suggests that the dominating runaway electrons emit most of their radiation at a wavelength $\lambdac\gg\SI{890}{nm}$. The relatively large vertical extent of the radiation also suggests that the dominating particles have pitch angles above $\thetap\sim\SI{0.25}{rad}$, an estimate that is arrived at through simulation of single-particle synchrotron radiation images.

\subsection{Kinetic modeling of the discharge}
\begin{figure}
    \centering
    \includegraphics[width=0.5\textwidth]{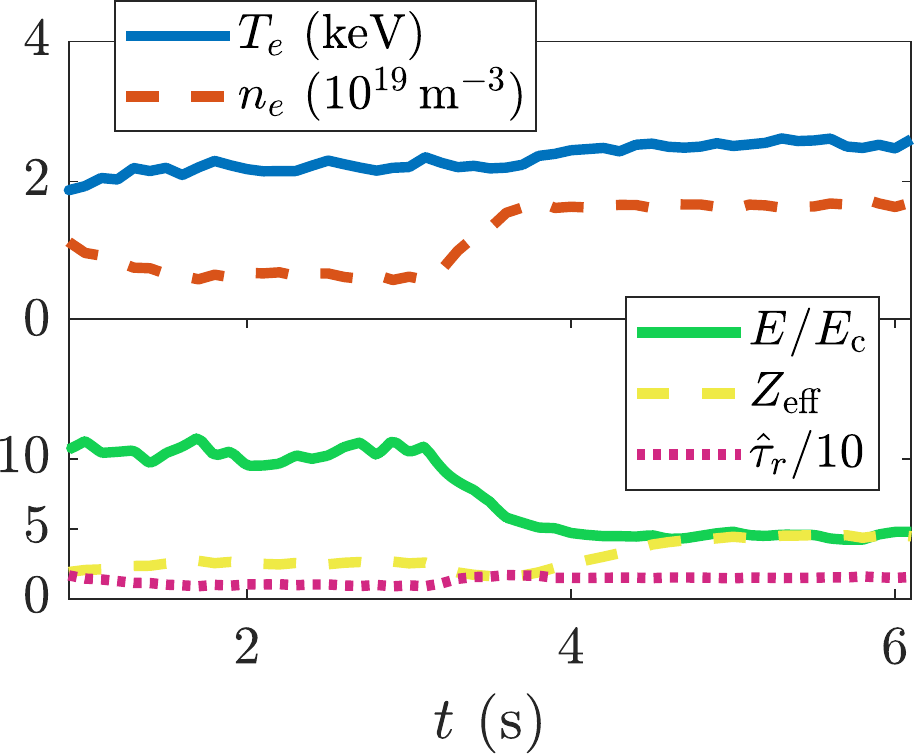}
    \caption{Temporal evolution during DIII-D discharge 165826 of the electron temperature $T_e$, electron density $n_e$ (upper plot), toroidal electric field normalized to the critical electric field $E_c$, the effective charge of the plasma $Z_{\mathrm{eff}}$ and ratio of the collision to synchrotron damping time $\hat{\tau}_r$ (lower plot).}
    \label{fig:profileEvolution}
\end{figure}

The temporal evolution of the 2D runaway electron momentum-space distribution function during DIII-D discharge 165826 was simulated using CODE~\cite{Landreman2014,Stahl2016} by solving the spatially homogeneous kinetic equation, including electric-field acceleration, collisions modeled by a linearized Fokker-Planck operator, avalanche source and synchrotron-radiation reaction losses. Temporal profiles of electron temperature, density, toroidal electric field and plasma effective charge used in the calculation are shown in Fig.~\ref{fig:profileEvolution}. All parameter profiles were measured at the magnetic axis, except for the electric field which was measured at the plasma edge. The electric field relaxation time is expected to be much shorter than the discharge time though, so that the radial profile of the electric field is expected to be approximately uniform.

\begin{figure}
    \centering
    \begin{overpic}[width=\textwidth]{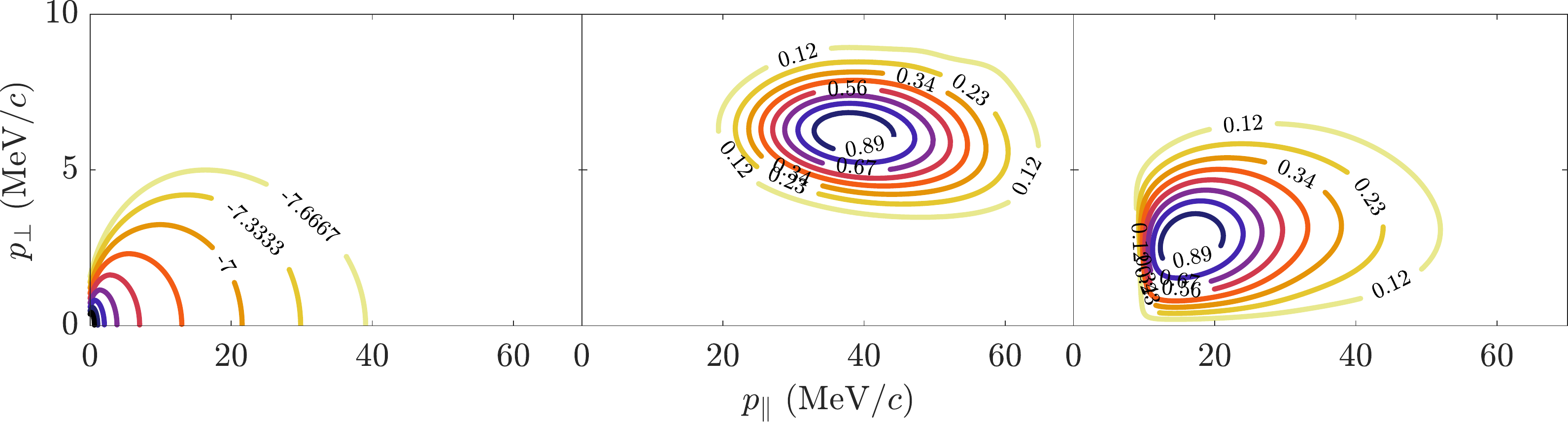}
        \put(6.5,22){\figlabel{(a)}}
        \put(12,22){\footnotesize{$\log_{10}\,f(p_\parallel, p_\perp)$}}
        \put(38,22){\figlabel{(b)}}
        \put(38,7.5){\figlabelsmall{Synchrotron emission}}
        \put(70,22){\figlabel{(c)}}
        \put(75,22){\figlabelsmall{Bremsstrahlung}}
    \end{overpic}
    \caption{Plot of \textbf{(a)} the distribution function, \textbf{(b)} the synchrotron radiation emission in momentum-space from the distribution function $F_s$ in the wavelength interval $\lambda\in[880,900]\,\mathrm{nm}$ and \textbf{(c)} the bremsstrahlung emission in momentum-space from the distribution function $F_b$ at photon energy $\SI{9}{MeV}$, with $F_s$ and $F_b$ defined in Eq.~\eqref{eq:radiationdensity}.}
    \label{fig:distfunc}
\end{figure}

In Fig.~\ref{fig:distfunc}, the resulting distribution function and corresponding synchrotron emission in momentum-space are shown. The relatively large number of runaway electrons with high energies causes the synchrotron emission to be dominated by runaway electrons with $\sim\SI{30}{MeV}$ energies and $\sim\SI{0.13}{rad}$ pitch angles. With bremsstrahlung we instead observe a different part of momentum-space, as the dominant runaways have energies around $\sim\SI{22}{MeV}$ and pitch-angles $\sim\SI{0.13}{rad}$, as illustrated in Fig.~\ref{fig:distfunc}(c).

\begin{figure}
    \centering
    \begin{overpic}[width=0.24\textwidth]{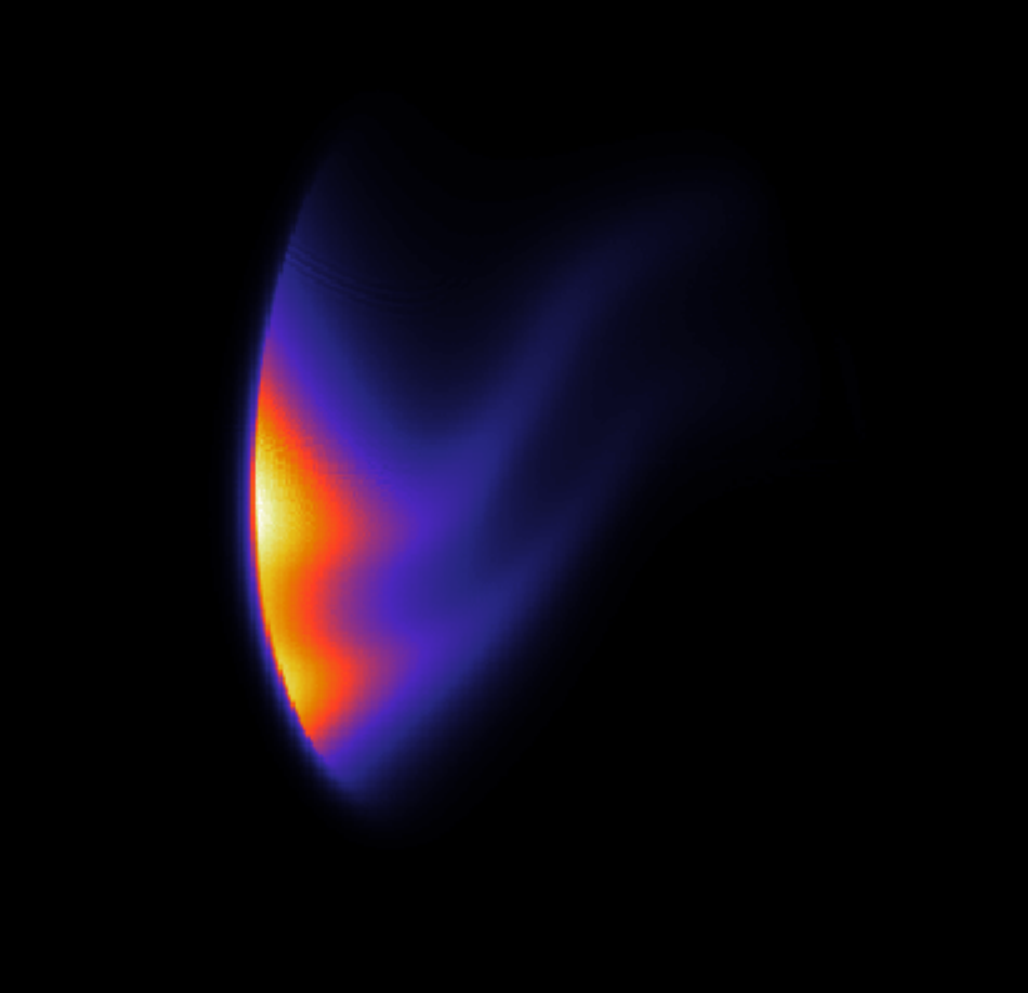}
        \put(5,82){\figlabelw{(a)}}
    \end{overpic}
    \begin{overpic}[width=0.24\textwidth]{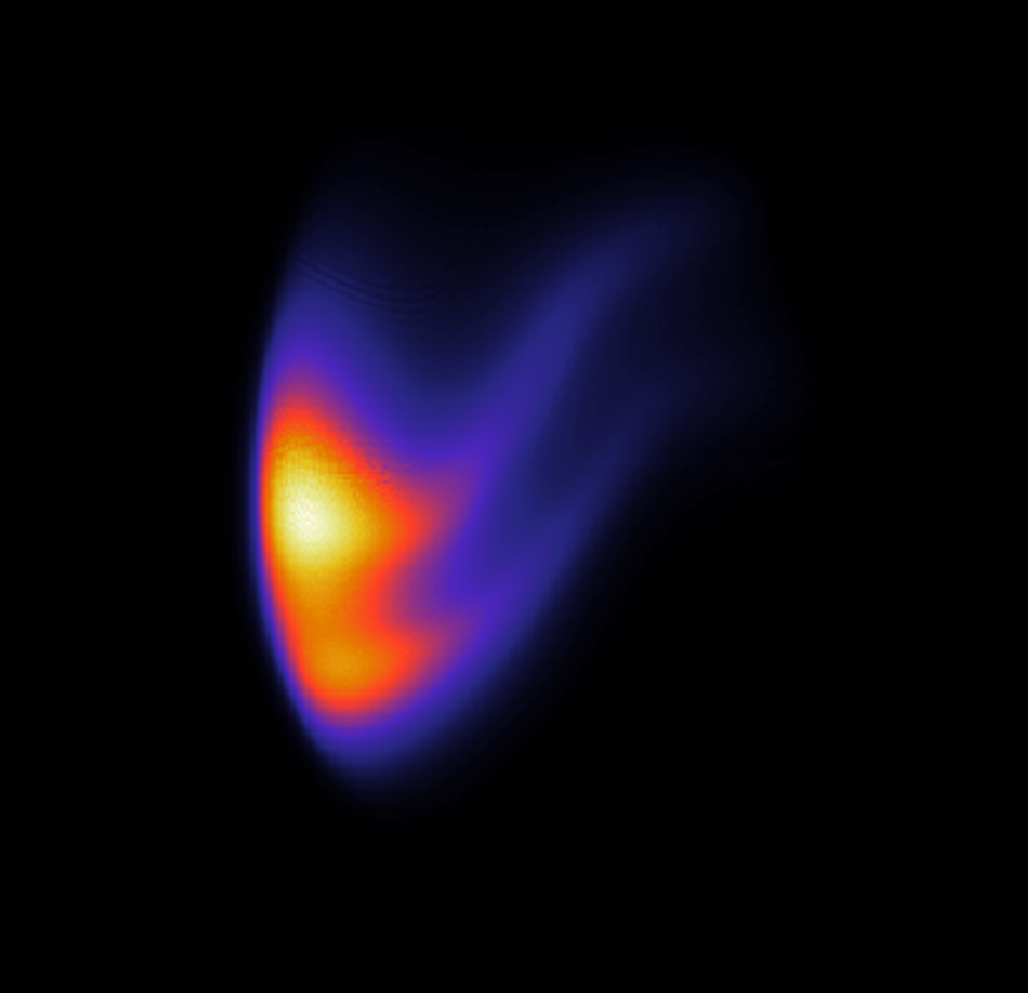}
        \put(5,82){\figlabelw{(b)}}
    \end{overpic}
    \begin{overpic}[width=0.24\textwidth]{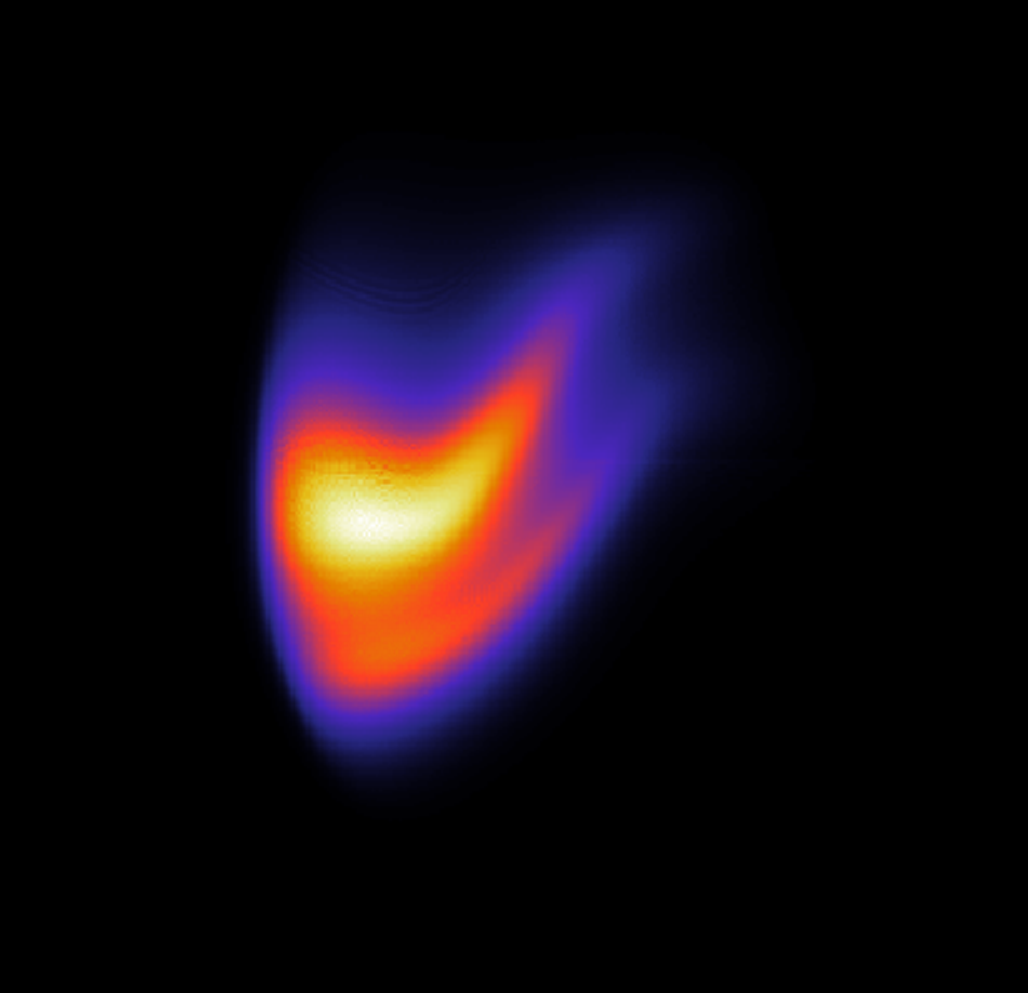}
        \put(5,82){\figlabelw{(c)}}
    \end{overpic}
    \begin{overpic}[width=0.24\textwidth]{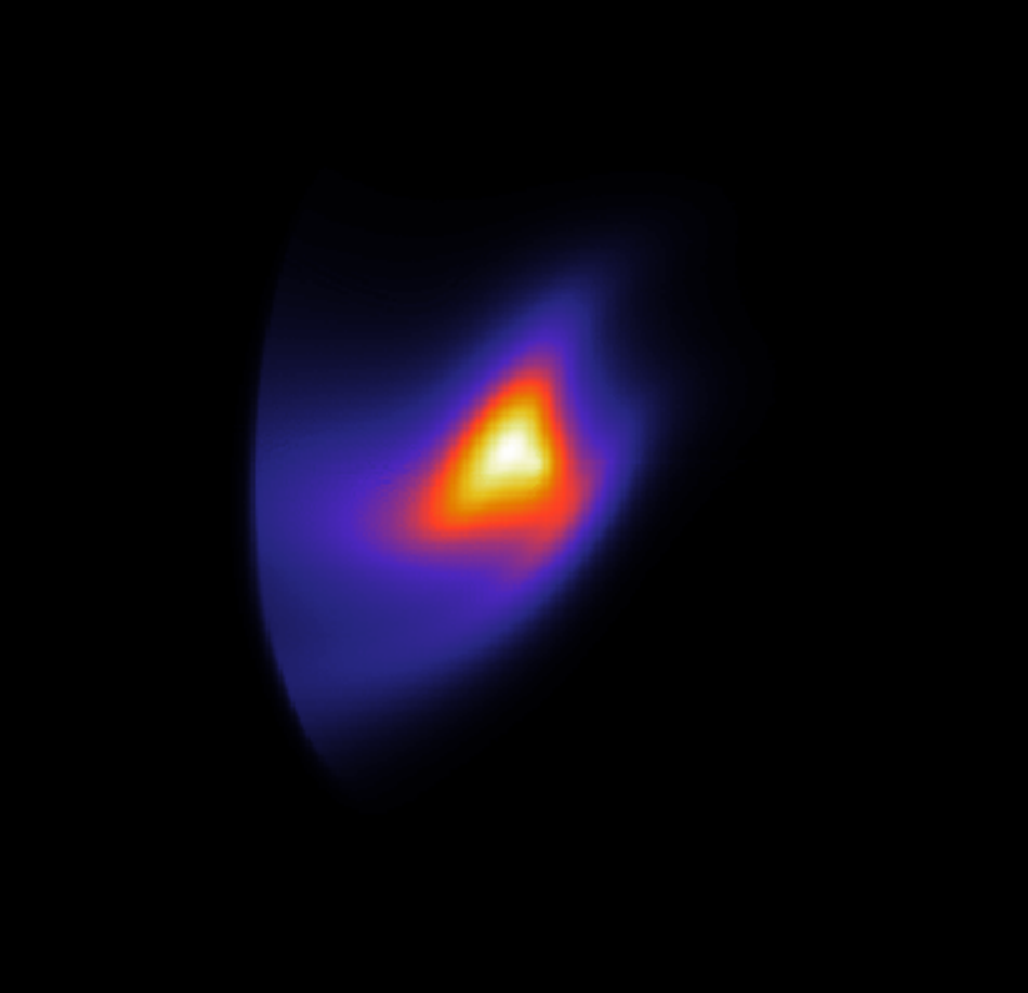}
        \put(5,82){\figlabelw{(d)}}
    \end{overpic}\\
    \begin{overpic}[width=0.24\textwidth]{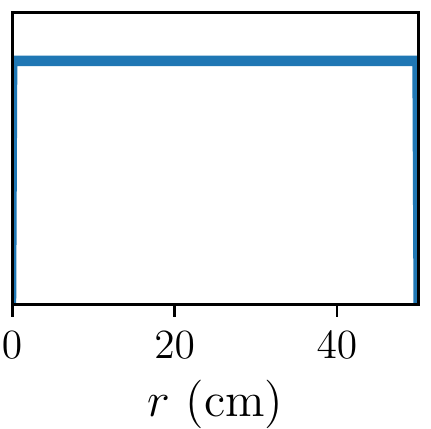}
        \put(8,37){\figlabel{(e)}}
    \end{overpic}
    \begin{overpic}[width=0.24\textwidth]{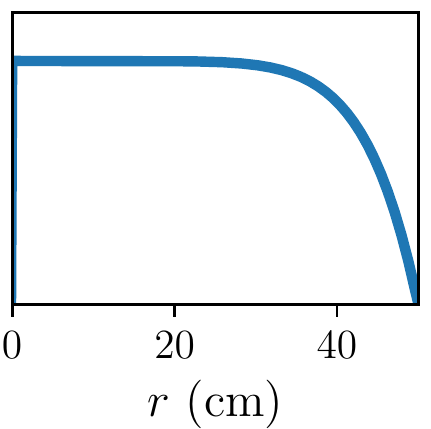}
        \put(8,37){\figlabel{(f)}}
    \end{overpic}
    \begin{overpic}[width=0.24\textwidth]{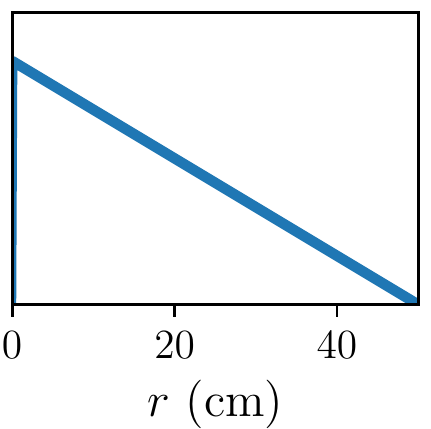}
        \put(8,37){\figlabel{(g)}}
    \end{overpic}
    \begin{overpic}[width=0.24\textwidth]{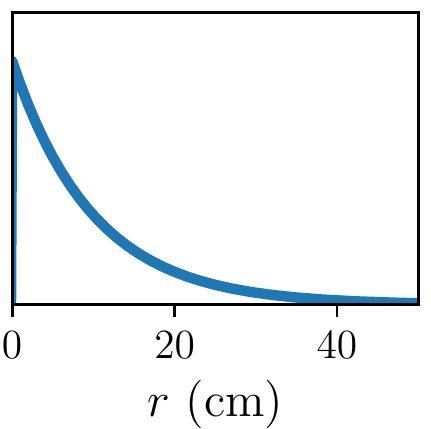}
        \put(8,37){\figlabel{(h)}}
    \end{overpic}
    \caption{Synthetic synchrotron image resulting from the simulated distribution function Fig.~\ref{fig:distfunc}(a). Four different radial profiles have been applied to this image to discern the possible shape of the actual radial profile. \textbf{(a/e)} Uniform/constant profile, cut off at $r = \SI{50}{cm}$. \textbf{(b/f)} Eighth-degree polynomial. \textbf{(c/g)} Linearly decreasing profile. \textbf{(d/h)} Exponentially decreasing profile. All profiles are plotted against minor radius, so that $r = 0$ corresponds to the magnetic axis.}
    \label{fig:synthimage}
\end{figure}

\subsection{Synchrotron radiation}
The synthetic synchrotron image resulting from the distribution function in Fig.~\ref{fig:distfunc} is shown in Fig.~\ref{fig:synthimage} with four different radial profiles applied to it, with the additional assumption that the momentum-space distribution is the same at all radii. As is seen in Fig.~\ref{fig:synthimage}, the radiation originates mainly from the HFS, which should be due to the finite spectral range effect described in Section~\ref{sec:spectrumeffect}. It is clear from Fig.~\ref{fig:synthimage} that none of the assumed radial profiles is suitable to reproduce all the details of the experimental image. Due to the smooth decrease in intensity towards the HFS in the experimental image Fig.~\ref{fig:synchcamera}, the radial profile should have to decrease smoothly to zero at larger radii, as in Figs.~\ref{fig:synthimage}(b)-(d).

The rather wide spot obtained with a linearly decreasing profile, and the significant contribution from near the magnetic axis with the exponentially decreasing profile, suggests that such a rapidly decreasing profile is unlikely to explain the experimental image, at least with the momentum-space distribution used here. 
An off-axis peak in the radial profile would provide even better agreement between simulation and experiment, however the two distinct, bright patches seen in the simulations would not go away with just a change in the radial distribution. Instead, the most plausible explanation is that the experimental runaway electron momentum-space distribution is different from the one predicted by solutions of the spatially homogeneous kinetic equation. There are important effects missing from this model, such as the effect of magnetic trapping, radial transport and drift-orbit losses that may be relevant to this DIII-D scenario. One further indication that it is the kinetic physics utilized that is not complete is, as mentioned, the appearance of two distinct, bright, vertically separated patches in the synthetic images. These bright patches should stem from the edges of the dominating SOV, and thus be the result of the line-integration effects described in Section~\ref{fig:sov}. The absence of these bright patches in experiment suggest that the experimentally observed SOV does not close on itself, which means that the dominating pitch angle must be much larger than the dominating pitch angle of the simulation.

The idea that kinetic processes not covered by the model employed are present in this DIII-D discharge was also suggested by \cite{PazSoldan2017}, where radial transport or a kinetic instability was given as possible explanations. The evidence for this provided by \cite{PazSoldan2017} was an energy distribution function inverted from measurements with the GRI diagnostic, which did not match the distribution function obtained from kinetic simulations. The inverted distribution function in \cite{PazSoldan2017} was characterized by much lower maximum energies than the corresponding simulated distribution function. This should make the difference between the high- and low-field side contributions more distinct in Fig.~\ref{fig:synthimage} and more consistent with Fig.~\ref{fig:synchcamera}.

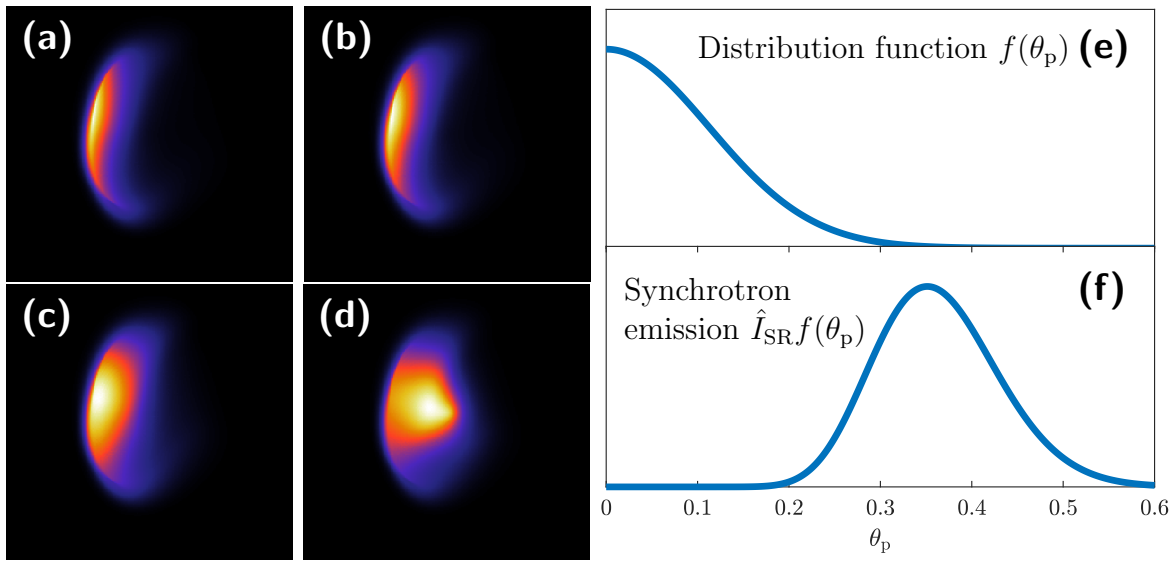
\begin{figure}
    \centering
    \begin{minipage}{0.49\textwidth}
        \begin{overpic}[width=0.49\textwidth]{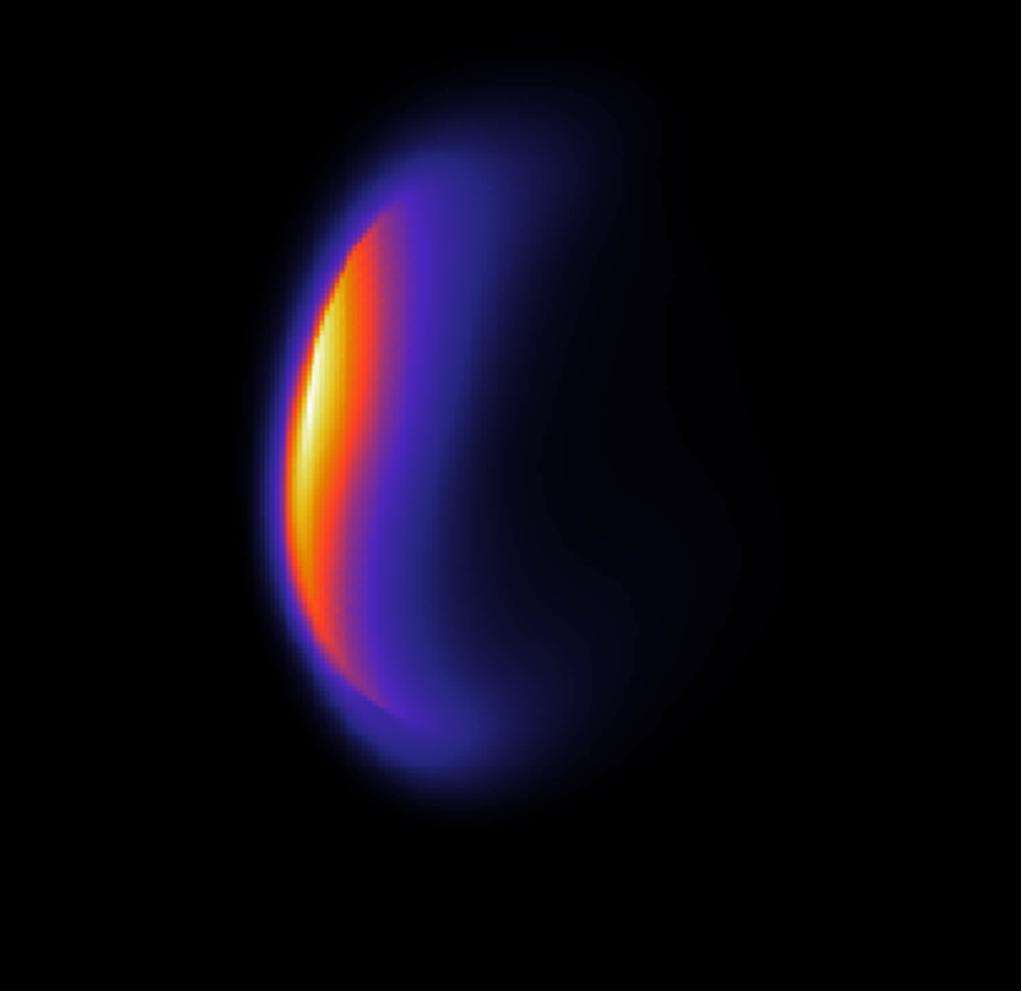}
            \put(5,82){\figlabelw{(a)}}
        \end{overpic}
        \begin{overpic}[width=0.49\textwidth]{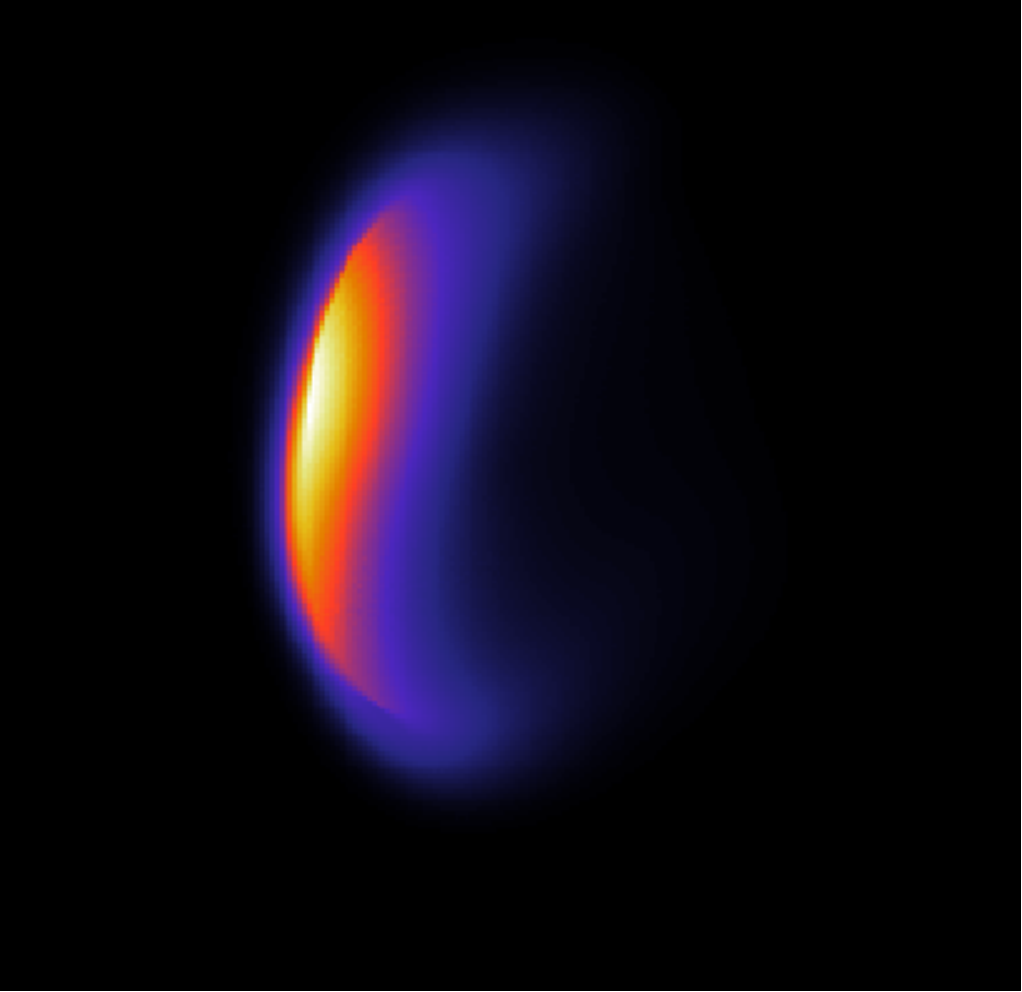}
            \put(5,82){\figlabelw{(b)}}
        \end{overpic}
        \begin{overpic}[width=0.49\textwidth]{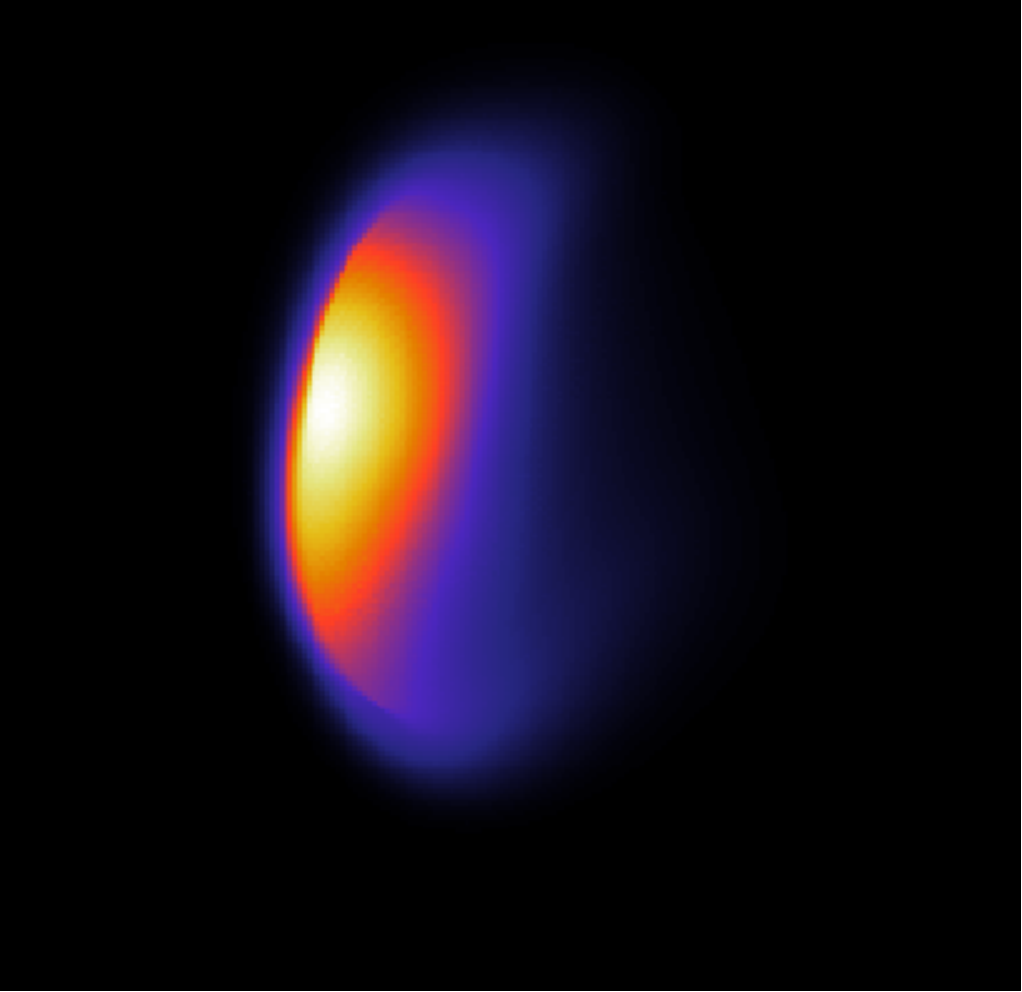}
            \put(5,82){\figlabelw{(c)}}
        \end{overpic}
        \begin{overpic}[width=0.49\textwidth]{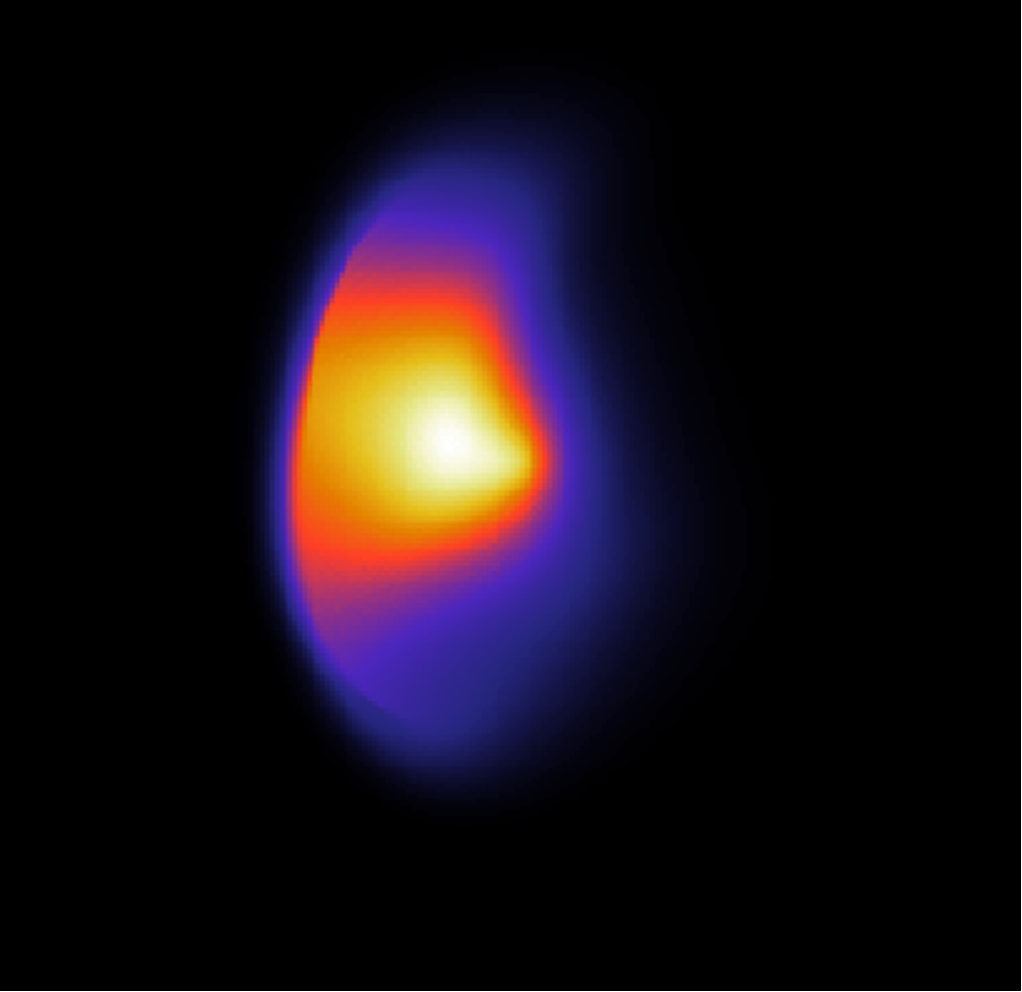}
            \put(5,82){\figlabelw{(d)}}
        \end{overpic}
    \end{minipage}
    \begin{minipage}{0.49\textwidth}
        \begin{overpic}[width=0.97\textwidth]{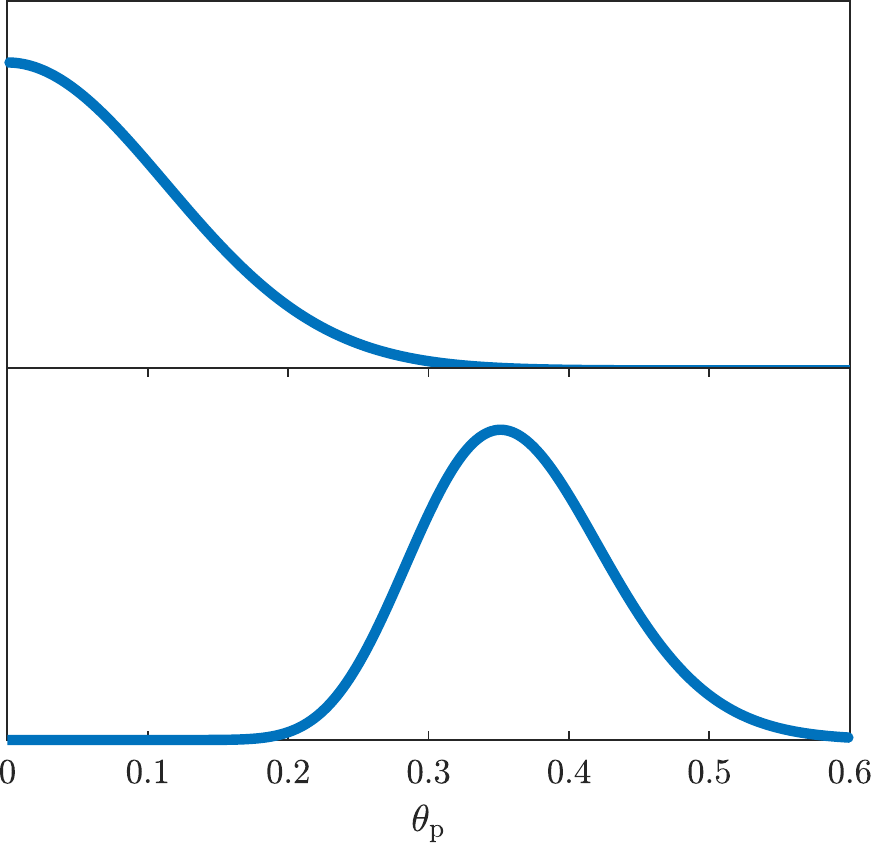}
            \put(84,88){\figlabel{(e)}}
            \put(17,88){Distribution function $f(\thetap)$}
            \put(84,45){\figlabel{(f)}}
            \put(4,45){Synchrotron}
            \put(4,38){emission $\hat{I}_{\mathrm{SR}}f(\thetap)$}
        \end{overpic}
    \end{minipage}
    \caption{Synchrotron images resulting from simulations with the toy distribution function shown in \textbf{(e)} with one runaway electron energy, $E = \SI{25}{MeV}$. In \textbf{(a)}-\textbf{(d)} the radial profiles Fig.~\ref{fig:synthimage}(e)-(h) have been applied. Part \textbf{(f)} shows the distribution function weighed with the synchrotron emission in pitch angle-space, which reveals that the dominating particle should have a pitch angle $\thetap\sim\SI{0.35}{rad}$.}
    \label{fig:toy}
\end{figure}

To test the hypothesis that Fig.~\ref{fig:synchcamera} is consistent with the dominating particle having a lower energy and larger pitch angle, a toy distribution function where all particles have the same energy $E = \SI{25}{MeV}$ but are distributed in pitch-angle so that the dominating particle has a pitch angle $\thetap\sim\SI{0.35}{rad}$. The distribution function is shown in Fig.~\ref{fig:toy}(e), and in Fig.~\ref{fig:toy}(f) the distribution function has been weighed with the synchrotron radiation weight function $\hat{I}_s$ (see Section~\ref{sec:distfunc}) to reveal from where in pitch-angle space that most synchrotron radiation will be emitted. In Figs.~\ref{fig:toy}(a)-(d) the resulting synthetic synchrotron image is shown, with the radial profiles of Fig.~\ref{fig:synthimage}(e)-(h) applied in order.

Of the synthetic images resulting from the toy distribution function in Fig.~\ref{fig:toy}, it is Figs.~\ref{fig:toy}(a) and (b) that most resemble the experimental image in Fig.~\ref{fig:toy}. They all have a crescent shape with only one bright patch, and in both Figs.~\ref{fig:toy}(a), (b) and Fig.~\ref{fig:synchcamera} the bottom end of the spot extends slightly further to the LFS than the upper end. This provides further evidence for the conclusion that additional kinetic effects beyond those included in the spatially-homogeneous model need to be invoked in order to understand the measurements. An additional energy-limiting mechanism could shift the distribution to lower energies, where the higher rate of pitch-angle scattering could plausibly produce the required shape of the runaway distribution. 

    \subsection{Bremsstrahlung}
\begin{figure}
    \centering
    \begin{overpic}[width=0.30\textwidth]{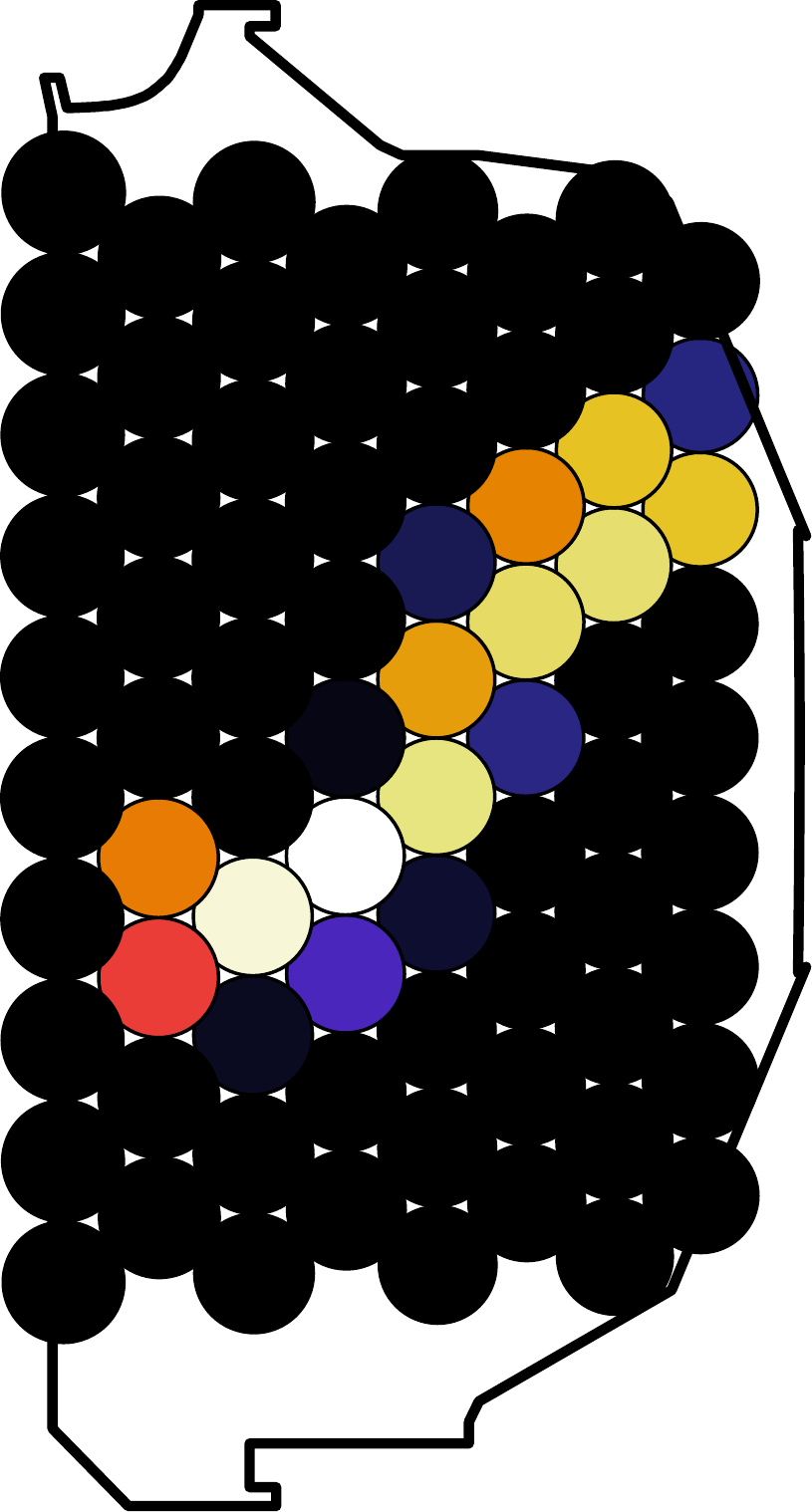}
        \put(0,100){\figlabel{(a)}}
        \put(24,100){$p_\parallel = \SI{20}{MeV/}c$}
        \put(24,95){$p_\perp = \SI{1}{MeV/}c$}
    \end{overpic}\hspace{8mm}
    \begin{overpic}[width=0.30\textwidth]{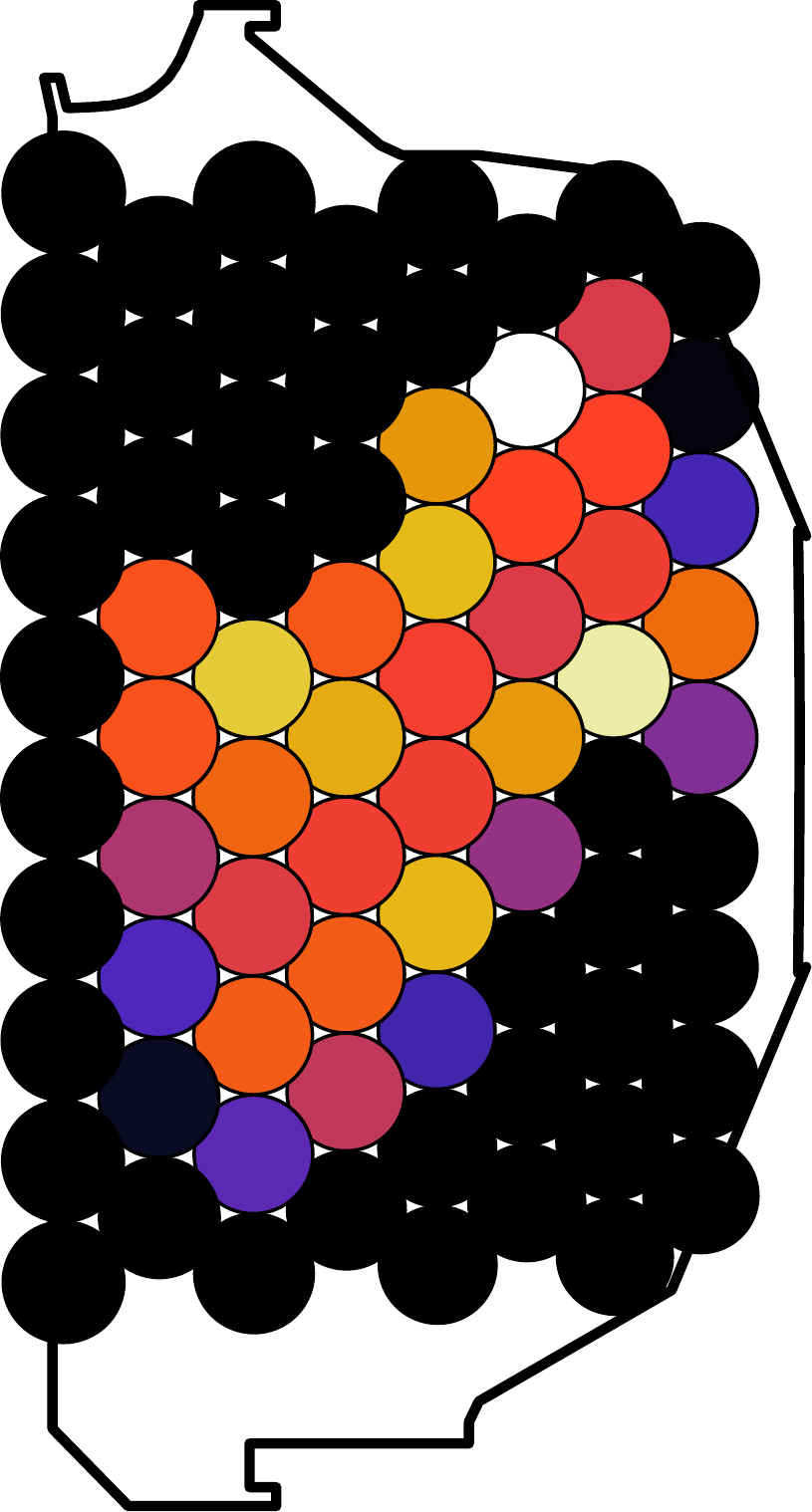}
        \put(0,100){\figlabel{(b)}}
        \put(24,100){$p_\parallel = \SI{20}{MeV/}c$}
        \put(24,95){$p_\perp = \SI{3}{MeV/}c$}
    \end{overpic}\hspace{8mm}
    \includegraphics[height=9cm]{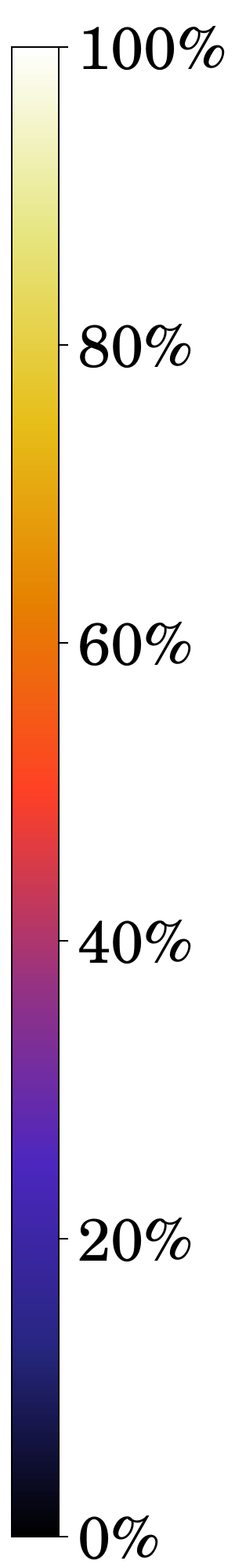}
    \caption{Synthetic GRI images resulting from particles with parallel momentum $p_\parallel = \SI{20}{MeV/}c$ and perpendicular momentum \textbf{(a)} $p_\perp = \SI{1}{MeV/}c$ and \textbf{(b)} $p_\perp = \SI{3}{MeV/}c$ respectively. The runaway electron beam radius was set to $\SI{50}{cm}$ and the detectors see photons in the range 1-$\SI{60}{MeV}$ uniformly.}
    \label{fig:bremssingle}
\end{figure}

\begin{figure}
    \centering
    \begin{overpic}[width=0.285\textwidth]{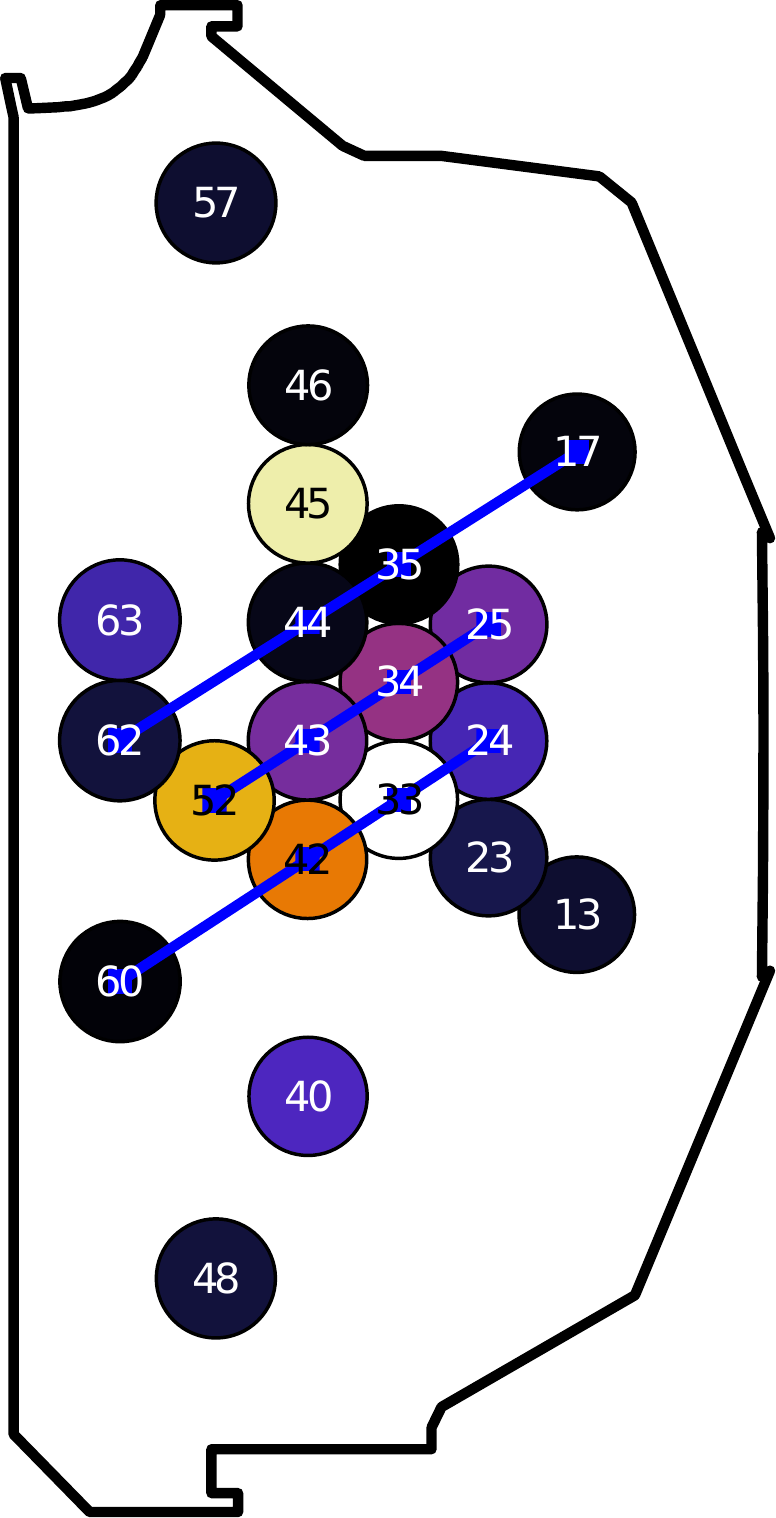}
        \put(0,100){\figlabel{(a)}}
        \put(24,100){\figlabelm{DIII-D}}
        \put(24,95){\figlabelm{\#165826}}
    \end{overpic}\hspace{8mm}
    \begin{overpic}[width=0.30\textwidth]{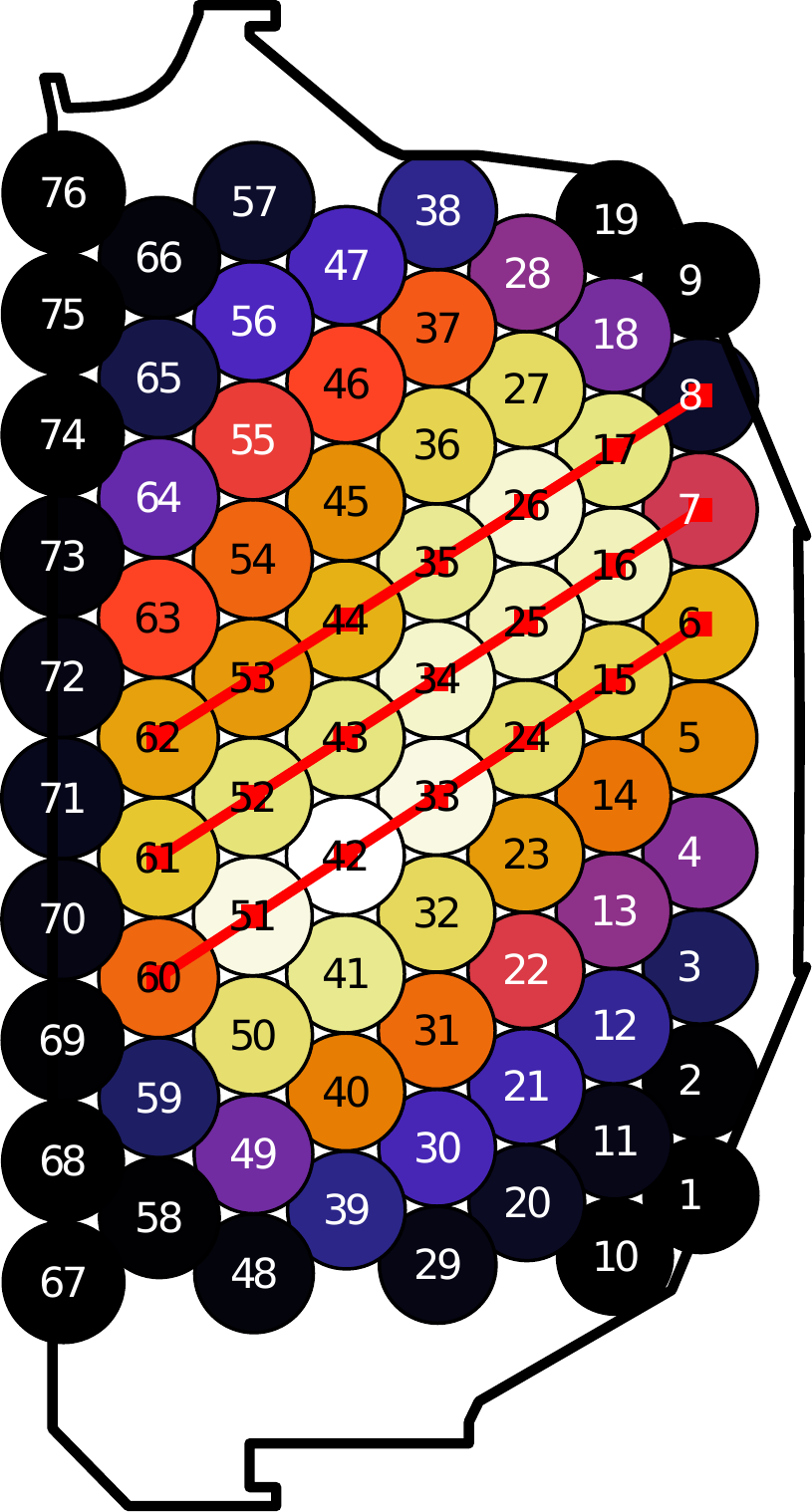}
        \put(0,100){\figlabel{(b)}}
        \put(24,100){\figlabelm{SOFT}}
    \end{overpic}\hspace{8mm}
    \includegraphics[height=9cm]{images/colorbar-vert.pdf}
    \caption{Comparison of \textbf{(a)} a GRI image reconstructed from experimental measurements to \textbf{(b)} a synthetic GRI image. The images show only radiation from the $\SI{9}{MeV}$ photon channel. All numbers correspond to detector indices and the red and blue lines mark the detectors that are plotted in Fig.~\ref{fig:gricomparison}.}
    \label{fig:bremsdist}
\end{figure}

In Figure~\ref{fig:bremssingle}, synthetic GRI images from mono-energetic and mono-pitch distributions simulated with SOFT are shown. Detectors are projected onto the poloidal plane orthogonal to the viewing direction of a central detector (no.\ 43 in Fig.~\ref{fig:bremsdist}, not considering the $z$ component of the viewing vector). In Fig.~\ref{fig:bremssingle}(a), a runaway electron beam with a $\SI{50}{cm}$ radius was initialized with parallel and perpendicular momentum $p_\parallel = \SI{20}{MeV/}c$ and $p_\perp = \SI{1}{MeV/}c$, while in Fig.~\ref{fig:bremssingle}(b) the particles were given $p_\parallel = \SI{20}{MeV/}c$ and $p_\perp = \SI{3}{MeV/}c$. The overall shape of the bremsstrahlung ``spot'' is the characteristic projection of a twisted cylinder, which results from the same SOV as synchrotron radiation, and in contrast to what we see in the synchrotron radiation images, the amount of radiation coming from the HFS and LFS seem to be roughly the same, due to the lack of a pitch angle dependence in the emitted bremsstrahlung spectrum Eq.~\eqref{eq:bremsspec}. The pitch angle does affect the spatial distribution of the bremsstrahlung though, since it determines the size and shape of the SOV. The effect of the runaway electron distribution function parameters on a bremsstrahlung image is therefore the same as shown for synchrotron radiation in Fig.~\ref{fig:spoteffects}, except for the camera finite spectral range effect in Fig.~\ref{fig:spoteffects}(b) which is completely absent.

\begin{figure}
    \centering
    \begin{overpic}[width=\textwidth]{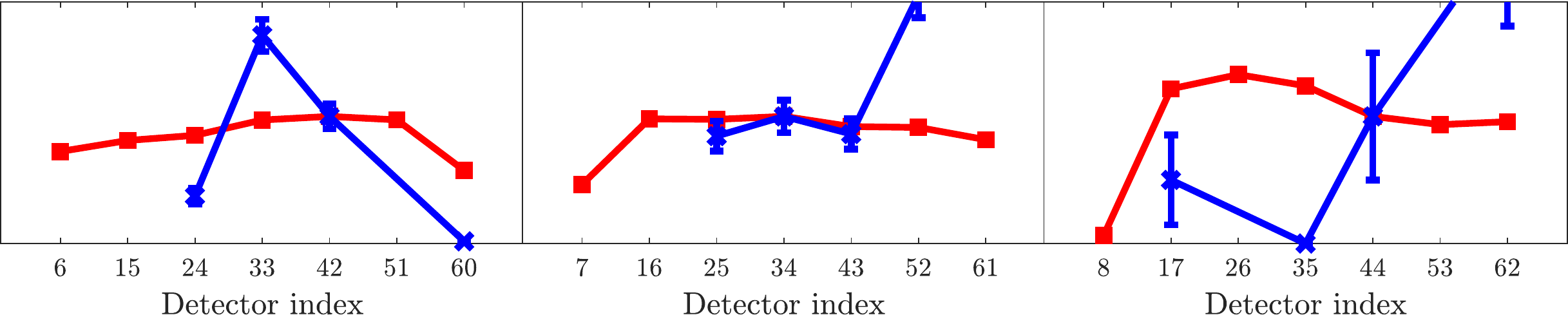}
        \put(1,16){\figlabel{(a)}}
        \put(34.5,16){\figlabel{(b)}}
        \put(68,16){\figlabel{(c)}}
    \end{overpic}
    \caption{Comparison of simulated detector signal (red squares) with the experimentally measured detector signals (blue crosses). The intensities have been normalized to the values of detector \textbf{(a)} 42, \textbf{(b)} 34 and \textbf{(c)} 44 to more clearly reveal similar trends in both datasets. Some detectors show similar trends, but several outlier points makes it difficult to draw any conclusions about the agreement between simulations and experiment.}
    \label{fig:gricomparison}
\end{figure}

The GRI was also used to determine the spatial distribution of bremsstrahlung during DIII-D discharge 165826, and the resulting measurement is shown in Fig.~\ref{fig:bremsdist}(a). In Fig.~\ref{fig:bremsdist}(b), the synthetic GRI image resulting from a SOFT simulation with the distribution function shown in Fig.~\ref{fig:distfunc}(a) is displayed. In Fig.~\ref{fig:gricomparison} the intensity variation in diagonally adjacent detectors is shown, and while some detectors show similar trends in simulation and experiment several outlier points make it difficult to draw any conclusions about agreement. A better comparison with experiment can hopefully be conducted in the future when more GRI sight-lines are populated with bremsstrahlung detectors.

\begin{figure}
    \centering
    \begin{overpic}[width=0.33\textwidth]{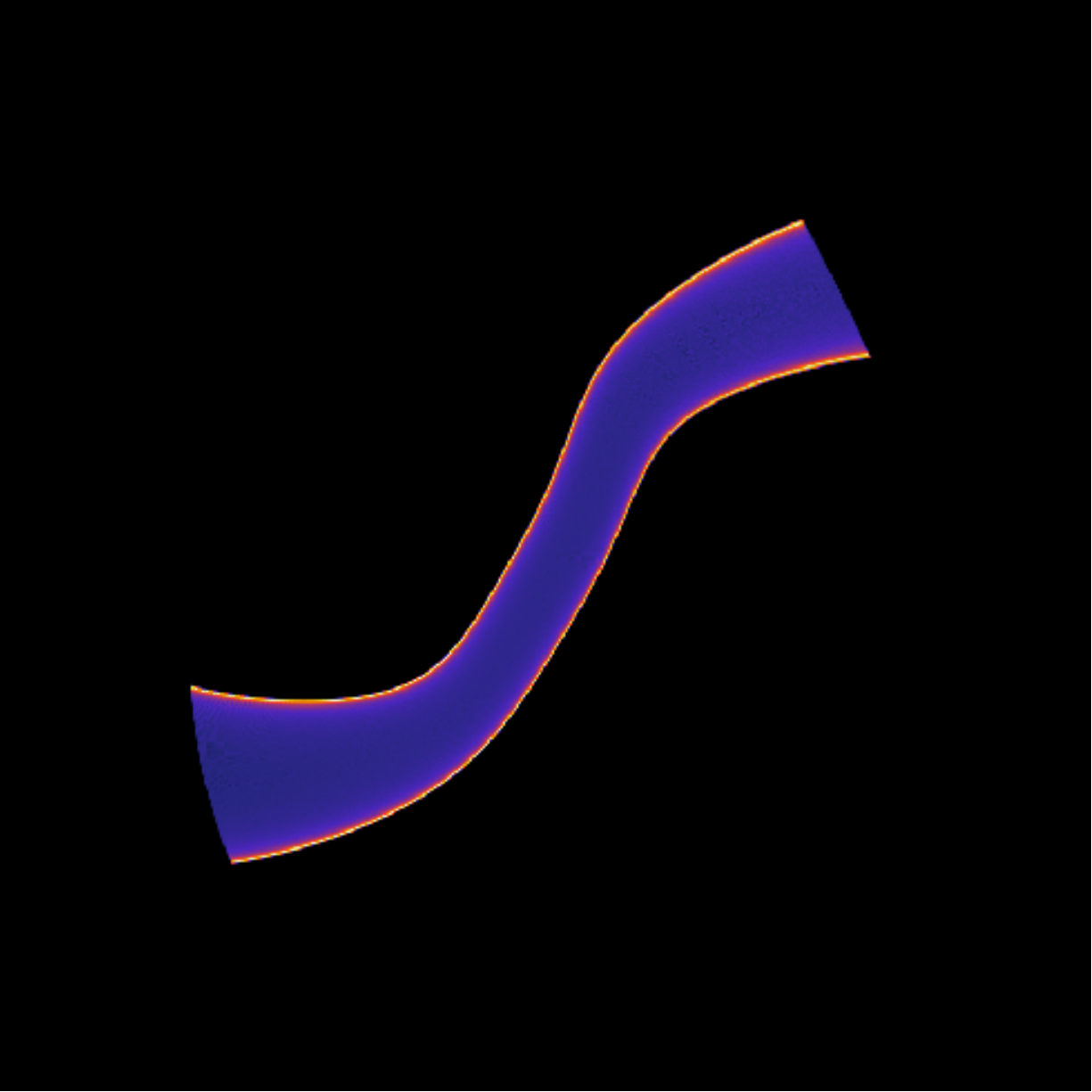}
        \put(5,82){\figlabelw{(a)}}
    \end{overpic}
    \begin{overpic}[width=0.33\textwidth]{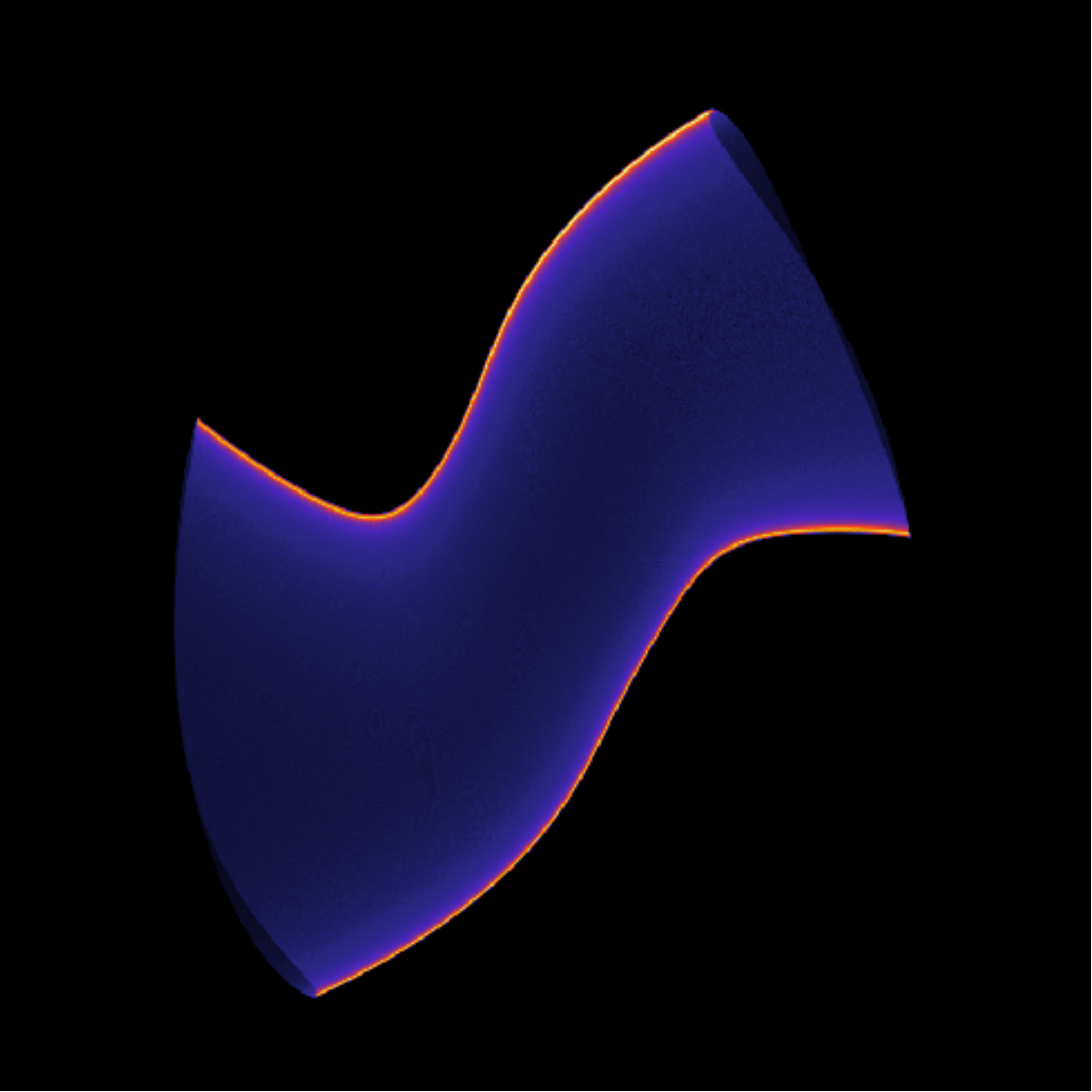}
        \put(5,82){\figlabelw{(b)}}
    \end{overpic}
    \begin{overpic}[width=0.33\textwidth]{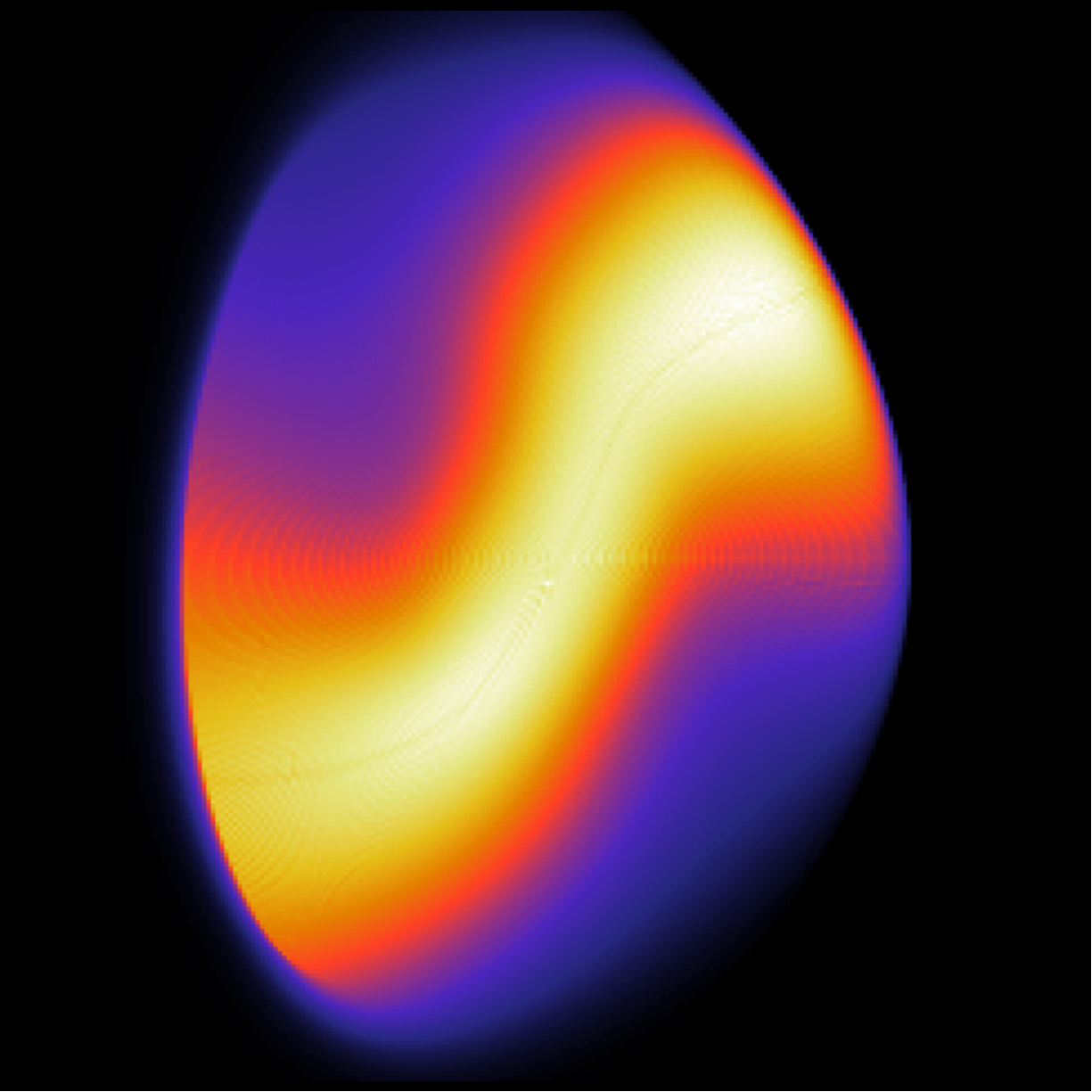}
        \put(5,82){\figlabelw{(c)}}
    \end{overpic}
    \caption{Bremsstrahlung images simulated with a synthetic high-resolution gamma-ray camera. Parts \textbf{(a)} and \textbf{(b)} result from single-particle simulations and correspond to the synthetic GRI images in Fig.~\ref{fig:bremssingle}, where $p_\parallel = \SI{20}{MeV}/c$ and \textbf{(a)} $p_\perp = \SI{1}{MeV}/c$ and \textbf{(b)} $p_\perp = \SI{3}{MeV}/c$. Part \textbf{(c)} results from a simulation with the distribution function in Fig.~\ref{fig:distfunc}(a), and corresponds to the GRI image in Fig.~\ref{fig:bremsdist}.}
    \label{fig:bremsideal}
\end{figure}

The relation between the bremsstrahlung and synchrotron radiation images becomes more apparent by using a more idealized synthetic camera for the bremsstrahlung simulation, as done in Fig.~\ref{fig:bremsideal}. Both types of radiation give rise to exactly the same SOV, with the same bright edges, and the radial dependence must therefore also be the same. Bremsstrahlung, in contrast to synchrotron radiation, has a more uniform radiation intensity distribution across the SOV since it is independent of magnetic field-strength, which means that bremsstrahlung is seen roughly equally strongly on both the HFS and LFS. Based on Fig.~\ref{fig:distfunc}(c), where the distribution function was weighed with the bremsstrahlung emission, we expect a wide range of pitch angles to contribute significantly to the emission. In the image, this appears as a very bright, S-shaped band in the center of the image, with gradually decreasing intensity in both vertical directions. This is the behavior seen in the simulated GRI image in Fig.~\ref{fig:bremsdist}(b), and becomes even more apparent in the high-resolution bremsstrahlung camera image Fig.~\ref{fig:bremsideal}(c).

    \section{Conclusions}\label{sec:conclusions}
    The synchrotron spot observed in runaway electron experiments is described in terms of the surface-of-visibility (SOV), the camera spectral range effect and the distribution function. The SOV determines the shape of the spot, while the camera spectral range determines the ratio of radiation seen on the HFS and LFS of the tokamak. The distribution function brings the individual synchrotron radiation spots from several classes of particles together and creates an overall spot shape, which combines the SOV and finite spectral range effect. Similar logic applies to bremsstrahlung, but due to the lack of pitch angle dependence in the bremsstrahlung emission, camera spectral range doesn't have the same dramatic effect on bremsstrahlung images, and its spatial localization depends only on the SOV.

An analysis of the formula for the synchrotron radiation spectrum in the limit of short wavelengths showed that the camera spectral range effect is very important in DIII-D, as the ratio between the intensity of synchrotron radiation on the HFS and LFS can be as high as $10^6$ using a visible light camera. In a synchrotron image this causes the radiation from the HFS part of the image to completely dominate, and should in most cases appear as a crescent.

Based on the synchrotron radiation simulations conducted in Section~\ref{sec:experiment}, we were able to conclude that the 0D2P linear Fokker-Planck simulations of the runaway electron distribution function did not satisfactorily reproduce the synchrotron measurements. If the dominant particle however had lower energy and associated larger pitch angle, the pattern was more accurately reproduced. The same conclusion was reached in Ref.~\cite{PazSoldan2017} for the same discharge and with a similar kinetic model, but based on bremsstrahlung measurements, reinforcing the suspicion of the kinetic model being insufficient. Effects not considered by our kinetic model that could play an important role in this DIII-D scenario include magnetic trapping, radial transport and drift-orbit losses. To further test this hypothesis, a 1D2P drift-kinetic solver such as LUKE~\cite{Peysson2003,Decker2004,Nilsson2015} or CQL3D~\cite{Chiu1998,Harvey2000} could be applied in order to also consider orbit effects.

Four different radial profiles were applied to the simulated synchrotron images and compared to the experimental image. A smooth decrease to zero at the runaway electron beam edge is necessary to reproduce the observed synchrotron spot, and the best matching radial profiles were those with a constant or almost constant behavior. There is great uncertainty about the shape of the radial profile closer to the magnetic axis however, since the finite camera spectral range effect hides all but the radiation furthest out on the HFS, so that no data points are available there.

Simulations of the GRI detector show similar trends between some individual detectors as the experimentally measured bremsstrahlung image, but the few data points available makes it difficult to draw any conclusions about agreement between simulations and experiment. The amount of information that could potentially be gained from GRI measurements in the future is however enormous, due to the availability of both spatial and spectral information from the diagnostic. The spatial information should be sufficient to constrain both the pitch angle and radial distribution functions, while the spectral data can be used to constrain the runaway electron energy. The independence on pitch angle and magnetic field strength of the bremsstrahlung emission also avoids the camera spectral range effect experienced with synchrotron radiation, which can conceal the distribution function near the magnetic axis. Thus the GRI is a promising diagnostic that, combined with simulations of the bremsstrahlung emission using SOFT, may be capable of finding more robust solutions to the inverse problem for the runaway electron distribution function.

    \ack
    The authors are grateful to L Hesslow, O Jakobsson, G Papp, I Pusztai and G Wilkie for fruitful discussions as well as the entire DIII-D team for excellent maintenance and operation of the tokamak. This work has been carried out within the framework of the EUROfusion Consortium and has received funding from the Euratom research and training programme 2014-2018 under grant agreement No 633053. The views and opinions expressed herein do not necessarily reflect those of the European Commission. The authors also acknowledge support from Vetenskapsr\aa det and the European Research Council (ERC-2014-CoG grant 647121). This material is based upon work supported by the U.S. Department of Energy, Office of Science, Office of Fusion Energy Sciences, using the DIII-D National Fusion Facility, a DOE Office of Science user facility, under Awards DE-FC02-04ER54698.
    
    \bibliographystyle{iopart-num}
    \bibliography{references}

\providecommand{\newblock}{}
\begin{thebibliography}{10}
\expandafter\ifx\csname url\endcsname\relax
  \def\url#1{{\tt #1}}\fi
\expandafter\ifx\csname urlprefix\endcsname\relax\def\urlprefix{URL }\fi
\providecommand{\eprint}[2][]{\url{#2}}

\bibitem{Dreicer1959}
Dreicer H 1959 {\em Physical Review\/} {\bf 115} 238

\bibitem{Jayakumar1993}
Jayakumar R, Fleischmann H~H and Zweben S~J 1993 {\em Physics Letters A\/} {\bf
  172} 447

\bibitem{RosenbluthPutvinski1997}
Rosenbluth M~N and Putvinski S~V 1997 {\em Nuclear Fusion\/} {\bf 37} 1355

\bibitem{hollmann15iter}
Hollmann E~M, Aleynikov P~B, F\"ul\"op T, Humphreys D~A, Izzo V~A, Lehnen M,
  Lukash V~E, Papp G, Pautasso G, Saint-Laurent F and Snipes J~A 2015 {\em
  Physics of Plasmas\/} {\bf 22} 021802

\bibitem{Hender2007}
Hender T, Wesley J, Bialek J, Bondeson A, Boozer A, Buttery R, Garofalo A,
  Goodman T, Granetz R, Gribov Y, Gruber O, Gryaznevich M, Giruzzi G, Günter
  S, Hayashi N, Helander P, Hegna C, Howell D, Humphreys D, Huysmans G, Hyatt
  A, Isayama A, Jardin S, Kawano Y, Kellman A, Kessel C, Koslowski H, Haye R~L,
  Lazzaro E, Liu Y, Lukash V, Manickam J, Medvedev S, Mertens V, Mirnov S,
  Nakamura Y, Navratil G, Okabayashi M, Ozeki T, Paccagnella R, Pautasso G,
  Porcelli F, Pustovitov V, Riccardo V, Sato M, Sauter O, Schaffer M, Shimada
  M, Sonato P, Strait E, Sugihara M, Takechi M, Turnbull A, Westerhof E, Whyte
  D, Yoshino R, Zohm H, the ITPA~MHD D and Group M~C~T 2007 {\em Nuclear
  Fusion\/} {\bf 47} S128

\bibitem{Boozer2017}
Boozer A~H 2017 {\em Nuclear Fusion\/} {\bf 57} 056018

\bibitem{Schwinger1949}
Schwinger J 1949 {\em Physical Review\/} {\bf 75} 1912

\bibitem{Bekefi}
Bekefi G 1966 {\em Radiation processes in plasmas\/} (New York: Wiley \& Sons)

\bibitem{Finken1990}
Finken K, Watkins J, Rusb{\"u}ldt D, Corbett W, Dippel K, Goebel D and Moyer R
  1990 {\em Nuclear Fusion\/} {\bf 30} 859

\bibitem{Jaspers1995}
Jaspers R, Grewe T, Finken K, Kr{\"a}mer-Flecken A, Cardozo N~L, Mank G and
  Waidmann G 1995 {\em Journal of Nuclear Materials\/} {\bf 220} 682

\bibitem{Jaspers2001}
Jaspers R, Lopes~Cardozo N~J, Donn{\'e} A~J~H, Widdershoven H~L~M and Finken
  K~H 2001 {\em Review of Scientific Instruments\/} {\bf 72} 466

\bibitem{Wongrach2014}
Wongrach K, Finken K, Abdullaev S, Koslowski R, Willi O, Zeng L and the
  TEXTOR~Team 2014 {\em Nuclear Fusion\/} {\bf 54} 043011

\bibitem{Esposito2003}
Esposito B, Mart\'in-Sol\'is J~R, Poli F~M, Mier J~A, S\'anchez R and
  Panaccione L 2003 {\em Physics of Plasmas\/} {\bf 10} 2350

\bibitem{Esposito2017}
Esposito B, Boncagni L, Buratti P, Carnevale D, Causa F, Gospodarczyk M,
  Martin-Solis J, Popovic Z, Agostini M, Apruzzese G, Bin W, Cianfarani C,
  Angelis R~D, Granucci G, Grosso A, Maddaluno G, Marocco D, Piergotti V, Pensa
  A, Podda S, Pucella G, Ramogida G, Rocchi G, Riva M, Sibio A, Sozzi C, Tilia
  B, Tudisco O, Valisa M and {FTU Team} 2017 {\em Plasma Physics and Controlled
  Fusion\/} {\bf 59} 014044

\bibitem{Tinguely2015}
Tinguely R~A, Granetz R~S, Stahl A and Mumgaard R 2015 {\em Bulletin of the
  American Physical Society\/} {\bf 60}

\bibitem{Papp2016IAEA}
Papp G, Pautasso G, Decker J, Gobbin M, McCarthy P~J, Choi D, Coda S, Duval B,
  Dux R, Erdos B {\em et~al.\/} 2016 Runaway electron generation and mitigation
  on the european medium sized tokamaks {ASDEX Upgrade and TCV} {\em
  Proceedings of the 2016 IAEA Fusion Energy Conference\/}
  \urlprefix\url{https://nucleus.iaea.org/sites/fusionportal/Shared%20Documents/FEC%202016/fec2016-preprints/preprint0502.pdf}

\bibitem{Vlainic2015}
Vlainic M, Mlynar J, Cavalier J, Weinzettl V, Paprok R, Imrisek M, Ficker O,
  Varavin M, Vondracek P and Noterdaeme J~M 2015 {\em Journal of Plasma
  Physics\/} {\bf 81} 475810506

\bibitem{Yu2013}
Yu J~H, Hollmann E~M, Commaux N, Eidietis N~W, Humphreys D~A, James A~N,
  Jernigan T~C and Moyer R~A 2013 {\em Physics of Plasmas\/} {\bf 20} 042113

\bibitem{Hollmann2013}
Hollmann E, Austin M, Boedo J, Brooks N, Commaux N, Eidietis N, Humphreys D,
  Izzo V, James A, Jernigan T, Loarte A, Martin-Solis J, Moyer R, Munoz-Burgos
  J, Parks P, Rudakov D, Strait E, Tsui C, Zeeland M~V, Wesley J and Yu J 2013
  {\em Nuclear Fusion\/} {\bf 53} 083004

\bibitem{PazSoldan2014}
Paz-Soldan C, Eidietis N~W, Granetz R, Hollmann E~M, Moyer R~A, Wesley J~C,
  Zhang J, Austin M~E, Crocker N~A, Wingen A {\em et~al.\/} 2014 {\em Physics
  of Plasmas\/} {\bf 21} 022514

\bibitem{Zhou2013}
Zhou R~J, Hu L~Q, Li E~Z, Xu M, Zhong G~Q, Xu L~Q, Lin S~Y, Zhang J~Z, {EAST
  Team} {\em et~al.\/} 2013 {\em Plasma Physics and Controlled Fusion\/} {\bf
  55} 055006

\bibitem{Cheon2016}
Cheon M, Kim J, An Y, Seo D and Kim H 2016 {\em Nuclear Fusion\/} {\bf 56}
  126004

\bibitem{Tong2016}
Tong R~H, Chen Z~Y, Zhang M, Huang D~W, Yan W and Zhuang G 2016 {\em Review of
  Scientific Instruments\/} {\bf 87} 11E113

\bibitem{Pankratov1996}
Pankratov I~M 1996 {\em Plasma Physics Reports\/} {\bf 22} 535

\bibitem{Pankratov1999}
Pankratov I~M 1999 {\em Plasma Physics Reports\/} {\bf 25} 145

\bibitem{Stahl2013}
Stahl A, Landreman M, Papp G, Hollmann E and F{\"u}l{\"o}p T 2013 {\em Physics
  of Plasmas\/} {\bf 20} 093302

\bibitem{Zhou2014}
Zhou R~J, Pankratov I~M, Hu L~Q, Xu M and Yang J~H 2014 {\em Physics of
  Plasmas\/} {\bf 21} 063302

\bibitem{Gomez2017KORC}
Carbajal L, del Castillo-Negrete D, Spong D, Seal S and Baylor L 2017 {\em
  Physics of Plasmas\/} {\bf 24} 042512

\bibitem{Gomez2017}
Carbajal L and del Castillo-Negrete D 2017 {\em Plasma Physics and Controlled
  Fusion\/} {\bf 59} 124001

\bibitem{Hoppe2017}
Hoppe M, Embr{\'e}us O, Tinguely R~A, Granetz R~S, Stahl A and F{\"u}l{\"o}p T
  2017 {\em To appear in Nucl. Fusion, arXiv preprint arXiv:1709.00674\/}

\bibitem{HoppeMSc}
Hoppe M 2017 {\em Synthetic synchrotron diagnostics for runaways in tokamaks\/}
  Master's thesis Chalmers University of Technology
  \urlprefix\url{http://publications.lib.chalmers.se/records/fulltext/249436/249436.pdf}

\bibitem{Pace2016}
Pace D~C, Cooper C~M, Taussig D, Eidietis N~W, Hollmann E~M, Riso V, {Van
  Zeeland} M~A and Watkins M 2016 {\em Review of Scientific Instruments\/} {\bf
  87} 043507

\bibitem{Cooper2016}
Cooper C~M, Pace D~C, Paz-Soldan C, Commaux N, Eidietis N~W, Hollmann E~M and
  Shiraki D 2016 {\em Review of Scientific Instruments\/} {\bf 87} 11E602

\bibitem{Koch1959}
Koch H~W and Motz J~W 1959 {\em Reviews of Modern Physics\/} {\bf 31} 920

\bibitem{PazSoldan2017}
Paz-Soldan C, Cooper C~M, Aleynikov P, Pace D~C, Eidietis N~W, Brennan D~P,
  Granetz R~S, Hollmann E~M, Liu C, Lvovskiy A, Moyer R~A and Shiraki D 2017
  {\em Phys. Rev. Lett.\/} {\bf 118}(25) 255002

\bibitem{Shevelev2017}
Shevelev A~E, Khilkevitch E, Lashkul S~I, Rozhdestvensky V~V, Pandya S~P,
  Plyusnin V~V, Altukhov A~B, Kouprienko D~V, Chugunov I, Doinikov D, Esipov
  L~A, Gin D, Iliasova M, Naidenov V, Polunovsky I, Sidorov A and Kiptily V~G
  2017 {\em Nuclear Fusion\/} Accepted for publication

\bibitem{Peysson2008}
Peysson Y and Decker J 2008 {\em Physics of Plasmas\/} {\bf 15} 092509

\bibitem{Nilsson2012}
Nilsson E, Decker J, Artaud J, Ekedahl A, Hillairet J, Peysson Y, Aniel T,
  Basiuk V, Goniche M, Imbeaux F {\em et~al.\/} 2012 Comparative modelling of
  lhcd with passive-active and fully-active multijunction launchers in the tore
  supra tokamak {\em Proc. 39th European Physical Society Conf. Plasma Physics
  and Controlled Fusion, Stockholm, Sweden\/}
  \urlprefix\url{http://ocs.ciemat.es/EPSICPP2012PAP/pdf/P2.081.pdf}

\bibitem{Landreman2014}
Landreman M, Stahl A and F{\"u}l{\"o}p T 2014 {\em Computer Physics
  Communications\/} {\bf 185} 847

\bibitem{Stahl2016}
Stahl A, Embr{\'e}us O, Papp G, Landreman M and F{\"u}l{\"o}p T 2016 {\em
  Nuclear Fusion\/} {\bf 56} 112009

\bibitem{Peysson2003}
Peysson Y, Decker J, Harvey R~W and Forest C~B 2003 Advanced {3-D} electron
  {Fokker-Planck} transport calculations {\em AIP Conference Proceedings\/} vol
  694 (AIP) p 495

\bibitem{Decker2004}
Decker J and Peysson Y 2004 {DKE}: {A} fast numerical solver for the {3D} drift
  kinetic equation Tech. Rep. EUR-CEA-FC-1736 Euratom-CEA

\bibitem{Nilsson2015}
Nilsson E, Decker J, Peysson Y, Granetz R, Saint-Laurent F and Vlainic M 2015
  {\em Plasma Phys. Controlled Fusion\/} {\bf 57} 095006

\bibitem{Chiu1998}
Chiu S, Rosenbluth M, Harvey R and Chan V 1998 {\em Nuclear Fusion\/} {\bf 38}
  1711

\bibitem{Harvey2000}
Harvey R, Chan V, Chiu S, Evans T, Rosenbluth M and Whyte D 2000 {\em Physics
  of Plasmas\/} {\bf 7} 4590--4599

\end{thebibliography}

\end{document}